\pdfoutput=1
\documentclass[]{jkpaper}
\usepackage{tikz}
\usetikzlibrary{decorations.markings}
\usetikzlibrary{decorations.pathreplacing}
\usetikzlibrary{arrows.meta}
\usetikzlibrary{patterns}
\usetikzlibrary{decorations.pathmorphing}

\usepackage{aas_macros}
\usepackage{soul}
\usepackage{accents}
\usepackage{simplewick}
\usepackage{cancel}

\newtheoremstyle{spaced}
{.8em}   
{.2em}   
{\itshape} 
{}        
{\bfseries} 
{.}       
{.5em}    
{}        

\theoremstyle{spaced}

\newcommand{\be}{
\begin{equation}}
  \newcommand{\ee}{
\end{equation}}
\newcommand{\baa}{
\begin{align}}
  \newcommand{\eaa}{
\end{align}}
\newcommand{\bea}{
\begin{eqnarray}}
  \newcommand{\eea}{
\end{eqnarray}}
\newcommand{\pa}{\partial}

\newcommand{\rd}{\mathrm{d}}
\newcommand{\onsheq}{\mathrel{\widehat{=}}}

\newcommand{\GN}{\ensuremath{G_{\text{N}}}}
\newcommand{\Aff}{\ensuremath{\operatorname{Aff}}}
\newcommand{\Diff}{\ensuremath{\operatorname{Diff}}}

\newcommand{\SLTR}{\ensuremath{\operatorname{SL}(2,\RR)}}
\newcommand{\Greens}[2]{\mathcal{G}(#1,#2)}

\makeatletter
\newcommand{\normord@delim}{%
  \mspace{1mu}%
  \vphantom{|}
  \smash{%
    \vcenter{%
      \offinterlineskip
      \hbox{\scalebox{0.55}{$\scriptscriptstyle\bullet$}}%
      \vskip .7ex
      \hbox{\scalebox{0.55}{$\scriptscriptstyle\bullet$}}%
    }%
  }%
  \mspace{1mu}%
}

\newcommand{\normord}[1]{%
  \vphantom{#1}\mathord{%
    \vphantom{|}
    \smash{%
      \mathinner{%
        \mathopen{\normord@delim}%
        \mkern1mu#1\mkern1mu%
        \mathclose{\normord@delim}%
      }%
    }%
  }%
}
\newcommand{\revnormord@delim}{%
  \mspace{1mu}%
  \vphantom{|}%
  \smash{%
    \vcenter{%
      \offinterlineskip
      \hbox{\scalebox{0.8}{$\scriptscriptstyle\boldsymbol{\times}$}}%
      \vskip .2ex
      \hbox{\scalebox{0.8}{$\scriptscriptstyle\boldsymbol{\times}$}}%
    }%
  }%
  \mspace{1mu}%
}

\newcommand{\revnormord}[1]{%
  \vphantom{#1}\mathord{%
    \vphantom{|}%
    \smash{%
      \mathinner{%
        \mathopen{\revnormord@delim}%
        \mkern0.6mu#1\mkern0.6mu%
        \mathclose{\revnormord@delim}%
      }%
    }%
  }%
}

\newcommand{\partcovnormord@delim}{%
  \mspace{1.5mu}%
  \vphantom{|}%
  \smash{%
    \vcenter{%
      \offinterlineskip
      \hbox{\scalebox{1.1}{$\scriptscriptstyle\star$}}%
      \vskip .35ex
      \hbox{\hspace*{0.21ex}\scalebox{0.55}{$\scriptscriptstyle\bullet$}}%
      \vskip .05ex
    }%
  }%
  \mspace{1.5mu}%
}

\newcommand{\partcovnormord}[1]{%
  \vphantom{#1}\mathord{%
    \vphantom{|}%
    \smash{%
      \mathinner{%
        \mathopen{\partcovnormord@delim}%
        #1%
        \mathclose{\partcovnormord@delim}%
      }%
    }%
  }%
}

\newcommand{\covnormord@delim}{%
  \mspace{1.5mu}%
  \vphantom{|}%
  \smash{%
    \vcenter{%
      \offinterlineskip
      \hbox{\scalebox{1.1}{$\scriptscriptstyle\star$}}%
      \vskip .35ex
      \hbox{\scalebox{1.1}{$\scriptscriptstyle\star$}}%
      \vskip .05ex
    }%
  }%
  \mspace{1.5mu}%
}

\newcommand{\covnormord}[1]{%
  \vphantom{#1}\mathord{%
    \vphantom{|}%
    \smash{%
      \mathinner{%
        \mathopen{\covnormord@delim}%
        #1%
        \mathclose{\covnormord@delim}%
      }%
    }%
  }%
}
\makeatother

\newcommand{\db}[2]{\pb{#1}{#2}_{\mathrm{D}}}

\usepackage{expl3,xparse}

\ExplSyntaxOn
\NewDocumentCommand{\Operator}{m}
{
  \tl_if_single:nTF { #1 }
  { \hat{#1} }
  { \widehat{#1} }
}

\NewDocumentCommand{\OperatorGaugeFixed}{m}
{
  \tl_if_single:nTF { #1 }
  { \hat{\bm{#1}} }
  { \widehat{\bm{#1}} }
}
\ExplSyntaxOff

\newcommand{\mv}{\mathrm{v}}

\makeatletter
\newcommand\footnoteref[1]{\protected@xdef\@thefnmark{\ref{#1}}\@footnotemark}
\makeatother

\titleformat{\paragraph}
{\normalfont\normalsize\bfseries}{\theparagraph}{1em}{}
\titlespacing*{\paragraph}
{0pt}{3.25ex plus 1ex minus .2ex}{1.5ex plus .2ex}

\newcommand{\Schwarzian}[2]{\left\{#1,#2\right\}}

\newcommand{\Ray}{\mathbb{R}}
\newcommand{\Segment}{\mathcal{I}}
\newcommand{\SegmentBar}{\bar{\Segment}}

\newcommand{\bg}[1]{\underline{#1}}

\newcommand{\Dmap}{\mathrm{D}}

\title{Localization and anomalous\texorpdfstring{\\}{ }reference frames
in gravity}
\author{Laurent Freidel\texorpdfstring{\textsuperscript{a}}{} and Josh Kirklin\texorpdfstring{\textsuperscript{b}}{}}
\institution{Perimeter Institute for Theoretical Physics,\texorpdfstring{\\}{ }31 Caroline Street North, Waterloo, ON, N2L 2Y5, Canada}

\email{
  \textsuperscript{a}\emaillink{lfreidel@pitp.ca}\\
  \textsuperscript{b}\emaillink{jkirklin@pitp.ca}
}

\abstr{
  In this work, we study the classical phase space for the gravitational degrees of freedom along a null ray. We construct gauge-invariant observables localized on a null ray segment that commute with those localized on the complement; thus, the phase space describes a genuine gravitational subsystem compatible with both locality and diffeomorphism invariance. Our construction employs `dressing time' (a null time coordinate built from spin 0 gravitational degrees of freedom) as a dynamical reference frame. The existence of such a frame depends on the use of edge mode variables, which we argue are generally required to upgrade a local gauge-fixing condition to a global `frame-fixing'. To analyze the effects of quantum diffeomorphism anomalies on these structures, we then establish an `effective' classical description in which the Raychaudhuri equation, symplectic form, and edge mode variables all acquire Virasoro-type deformations. Within this framework, we identify three distinct diffeomorphism actions: reparametrizations (gauge transformations), reorientations (physical symmetries of the reference frame), and dressed reparametrizations. Each acquires its own central extension and plays a different crucial role in the effective theory. The resulting structures provide a foundation for quantizing gravitational null ray segments, including promoting dressing time to a genuine quantum reference frame.
}

\begin{document}
\maketitleandtoc

\section{Introduction}
\label{Section: Introduction}

How should we reconcile gravitational diffeomorphism invariance with the primary role of local experiments and observations in physics? This question lacks a simple answer because truly local diffeomorphism-invariant observables do not exist~\cite{Torre1993}. Indeed, any non-trivial observable supported at a fixed spacetime point (or supported on a fixed proper subregion of spacetime) cannot be diffeomorphism-invariant, since diffeomorphisms move the support. For this reason, locality in gravity cannot meaningfully be attached to a fixed background structure. Rather, one must adopt a notion of \emph{relational locality}, in which observations are performed at spacetime locations determined by physical reference degrees of freedom~\cite{Bergmann1960,Bergmann1961a,Bergmann1961, Rovelli1991,Isham1993, Brown:1994py, Rovelli2002, Dittrich2006, Dittrich2007a, Dittrich2007, Khavkine2015, Kuchar2011,Goeller:2022rsx}.

Many mysteries of quantum gravity can be attributed to the fact that these degrees of freedom must themselves ultimately be quantum. They comprise a \emph{quantum reference frame} (QRF), and accounting for the necessity of such QRFs simultaneously enables a consistent picture of local gravitational physics, and fundamentally modifies it -- because any local observations are inextricably tied up with the quantum properties of the frame~\cite{Isham1985,Giddings_2006,Marolf_2015,Donnelly:2016rvo,Donnelly2016,brunetti2015quantumgravitypointview,Gambini2001,Husain_2012,Gary:2006mw,Baldazzi2022,Dittrich_2017,Donnelly2017,Giddings_2019,CastroRuiz2020,Giddings:2025xym,Kuchar2011,Thiemann2006,isham1995structuralissuesquantumgravity,Giesel_2010,Giesel:2020raf,Giesel:2024xtb}.

Quantum mechanical QRFs like simple clocks are by now relatively well understood (see~\cite{Hoehn2019,Hoehn2020,Hoehn2021,Hoehn2021a,Hoehn2022,Hoehn2023,Giacomini2019,Giacomini2021,Angelo2011,Loveridge2019,Hamette2020,Hamette2021} for some approaches), and have for example led to much recent progress in our understanding of gravitational entropies~\cite{CLPW,Jensen2023,Fewster2025,DeVuyst:2024pop,DeVuyst:2024uvd,kirklin2024generalisedsecondlawsemiclassical,KudlerFlam2025,Klinger2024,Klinger2024,KudlerFlam2025a,Faulkner2024}. But such QRFs are only sensitive to finite-dimensional subgroups of the diffeomorphism group, such as the one-dimensional flow corresponding to a particular notion of time evolution. Far richer structures should emerge once we consider more fully-fledged gravitational QRFs, i.e.\ quantum coordinate systems sensitive to genuine diffeomorphism groups. Such QRFs must be made of quantum \emph{fields}, and QRFs of this kind are rather less well understood than their quantum mechanical counterparts (although various different approaches have been taken, see for example~\cite{Hoehn2024,fewster2020quantumfieldslocalmeasurements, Gary:2006mw,kaplan2025sitterquantumgravityemergence,Kolchmeyer:2024fly,Kabel2023,Chen2024,Fewster:2025ijg}). One of the main difficulties is that in gauge theories the QRFs needed are infinite dimensional fields, which leads to new challenges. Some of these challenges have started to be tackled in the study of asymptotic symmetries, where the QRFs there are identified as Goldstone modes for the broken asymptotic symmetries 
\cite{Kapec:2017tkm, Nande:2017dba, Arkani-Hamed:2020gyp, Himwich:2020rro, Donnay:2022hkf, Agrawal:2023zea}. 

In gravity, the  reference frames can be classically described as  `embedding fields', which are maps $X$ from some reference coordinate space into spacetime~\cite{Donnelly:2016auv, Freidel:2020ayo,Freidel:2020svx, Freidel:2021dxw, Ciambelli:2021nmv,Carrozza2022,Carrozza2024,Goeller:2022rsx,Freidel:2020xyx,}.
Unlike ordinary fixed coordinate systems, embedding fields $X$ are dynamical objects, and transform covariantly under diffeomorphisms. Hence, by pulling back spacetime fields $\phi$ through the embedding map, one obtains gauge-invariant `dressed fields' $\tilde\phi:=X^*\phi$ which are relationally local to the coordinates provided by $X$~\cite{Goeller:2022rsx, Ciambelli:2023mir, Ciambelli_2024}. One can then define a gauge-invariant gravitational subsystem as the image under $X$ of a subregion in the reference space. Doing this generically necessitates\footnote{A notable exception is when the gravitational system includes dust fields that can act as gravitational QRFs \cite{Brown:1994py, Giesel_2010, Giesel:2012rb}. In this case, no edge mode is needed to achieve localization. This is however, a non-generic case.  } the inclusion of edge modes at the boundary of the subregion, and physical `corner symmetries' acting on the edge modes~\cite{Donnelly:2016auv,Geiller:2017xad,Geiller:2019bti,Ciambelli:2021nmv,Ciambelli:2021vnn, Freidel:2023bnj,}. One way to understand the edge modes is that they provide boundary conditions required for integrating a local gauge-fixing condition into a \emph{complete} gauge-fixing, or `frame-fixing'; we shall comment on this more in the paper. Note that part of the gravitational edge modes is a specification of the location of the boundary of the subregion, which is necessary for even discussing the subregion in the first place. In~\cite{kirklin2024generalisedsecondlawsemiclassical}, it was shown that these location edge modes' free energy contributes via non-semiclassical corrections to the Generalized Second Law.

It is important to note that the edge modes can come from different places, depending on the setup. Indeed, they can be `internal' \cite{Ciambelli:2021nmv, Freidel:2021dxw}, meaning in this paper that they are part of the degrees of freedom inside the subregion, or `external', meaning they are part of the degrees of freedom outside the subregion. In fact, it is only the \emph{differences} between internal and external edge modes which are gauge-invariant, and thus represent physical degrees of freedom~\cite{Araujo-Regado:2024dpr}; this will be important for the present paper, in which we will be concerned with the relative boosts between internal and external embedding fields.\footnote{Another important distinction concerns whether the edge modes are comprised of dynamical degrees of freedom already present in the phase space, or auxiliary degrees of freedom. Sometimes the former are called `internal' and the latter `external', but we use this language to refer to the inside and outside the subregion, as described in the main text. We thank Julian De Vuyst, Philipp H\"ohn, Luca Marchetti, Gon\c{c}alo Araujo Regado, Francesco Sartini, and Bilyana Tomova for clarifying these distinctions for us. }

We will focus on a particular gravitational reference frame called the `dressing time'~\cite{Ciambelli:2023mir,Ciambelli_2024}, which is a particular null time coordinate along a gravitational null surface (as such it is the inverse of the embedding field, which we denote $V=X^{-1}$). For various reasons, null surfaces provide the simplest setting in which both relational localization and gravitational dressing can be implemented explicitly. The first remarkable fact is that the projection of the gravitational constraints on a null surface can be exactly solved and constraint-free initial data can be found~\cite{Sachs:1962zzb, Torre:1985rw, Gourgoulhon_2006,Reisenberger:2007pq,Reisenberger:2012zq,Reisenberger:2018xkn,Mars:2022gsa, Freidel:2024emv,ciambelli2025foundationscarrolliangeometry}. Moreover, in recent years, the detailed construction of the gravitational symplectic structure on null surfaces has been achieved in~\cite{Parattu:2015gga, Lehner:2016vdi, Hopfmuller:2016scf, DePaoli:2017sar, Wieland:2017zkf, Hopfmuller:2018fni, Chandrasekaran:2018aop, Donnay:2019jiz, Chandrasekaran2021, Adami:2021nnf, Sheikh-Jabbari:2022mqi, Chandrasekaran:2021hxc, Adami:2021nnf, Sheikh-Jabbari:2022mqi, Ciambelli:2023mir,Chandrasekaran:2023vzb, ,Odak:2023pga, Rignon-Bret:2024wlu, Wieland:2024dop}.
The main results synthesized in  \cite{Ciambelli:2023mir} are that, on a null surface, the gravitational degrees of freedom naturally split into spin 0 modes associated with the area element, spin 1 modes associated with the null generators of the surface, and spin 2 modes associated with gravitational radiation. 
Finally, the principle of ultralocality (i.e., the fact that different null rays on the surface are causally disconnected) \cite{kay1991theorems, Schroer:2010bjq, Jensen2023, Wall:2011hj,Ciambelli_2024} allows us to analyze each null ray individually. 

The dynamics are then governed by the null-null component of the Einstein equation, i.e.\ the Raychaudhuri equation, which in the gravitational phase space acts as a constraint generating diffeomorphisms along the ray. Subsystems of the null ray are \emph{null ray segments}, and the corner symmetry group for such segments reduces to the affine group (containing translations and boosts of the segment endpoints). The dressing time is a natural `preferred' reference frame for these null ray segments, made entirely from spin 0 data. It has many convenient properties (for example, it commutes with itself in the decoupling regime described below, and is the natural time coordinate in which to express the second law~\cite{Ciambelli:2023mir,Ciambelli_2024, Wieland:2024dop}), which are not satisfied by other common choices of null time coordinate (such as affine time).

The Raychaudhuri constraint involves interactions between the spin 0 degrees of freedom and the rest of the fields. Thus, the reference frame (the dressing time) is coupled with the system it is used to observe (the other fields). In general, interactions of this kind can complicate the analysis of reference frames and the observables dressed to them (see for example~\cite{Dittrich_2017, dittrich2015chaosdiracobservablesconstraint,Paiva_2022}). To avoid these complications, we shall make a certain change of variables and go to a simplified regime in which the spin 2 gravitational radiation is perturbative, and any matter is just a collection of free scalar fields. This allows us to neglect any interactions between the frame and system. More precisely, we put the total Raychaudhuri constraint $C$ in the non-interacting form $C=H_R+H_S$. Here, $H_R$ is the stress contribution of the dressing time frame degrees of freedom, while $H_S$ is the stress tensor for matter and gravitational  radiation variables, which comprise the system (we explain this in more detail below). This kind of decoupled constraint is typical of much of the QRF literature, including recent developments to do with the algebraic crossed product \cite{Williams2007CrossedPO} and gravitational entropies~\cite{CLPW,Jensen2023,Fewster2025,DeVuyst:2024pop,DeVuyst:2024uvd,kirklin2024generalisedsecondlawsemiclassical,KudlerFlam2025,Klinger2024,Klinger2024,KudlerFlam2025a,Faulkner2024}.

The purpose of this paper is threefold:

\begin{itemize}

  \item First, we construct an explicit classical phase space describing the localized gravitational degrees of freedom on a null ray segment, in which observables supported on the segment commute with those supported on its complement. This provides a concrete realization of a gravitational subsystem compatible with locality, causality, and diffeomorphism invariance. The null ray segment is relationally local with respect to the `dressing time' reference frame~\cite{Ciambelli:2023mir}. The definition of this reference frame requires one to use edge mode degrees of freedom~\cite{Donnelly:2016auv, Freidel:2023bnj,Geiller:2017xad,Geiller:2019bti,Law:2025ktz,Ciambelli:2021nmv,Ciambelli:2021vnn} which come from outside the segment. We will not need to depend on a particular prescription for their exact origin. 

  \item Second, we point out the importance of three kinds of diffeomorphism symmetries: reparametrizations, dressed reparametrizations, and reorientations.
The reparametrizations and dressed reparametrizations are gauge transformations associated with constraints, while reorientations are physical symmetries possessing a non-vanishing charge.
The reparametrizations are gauge transformations acting on the fields defined with respect to background coordinates, while the dressed versions are analogous transformations of the dressed gauge-invariant fields. 
The reorientations, on the other hand, change the dressing time without affecting the system variables; in other words, they only change the `orientation' of the reference frame. They correspond to the original edge mode symmetry of \cite{Donnelly:2016auv} and the name reorientation was chosen to match the QRF literature~\cite{Hamette2021,Carrozza2022,Carrozza2024}. 
We show that the generator of reorientations is the second derivative of the gauge-invariant dressed area element along the null ray.
  
  \item Third, we show that the null ray segment is still a well-defined local subsystem in the presence of diffeomorphism anomalies. Indeed, such anomalies represent one of the main difficulties in constructing QRFs from fields and are expected to have deep consequences for the structure of the quantum theory, including notions of locality, subsystems, and edge modes. We will analyze these consequences for the properties of the local degrees of freedom on the null ray segment. 
  Despite our motivations being quantum, we shall work entirely at the classical level, i.e.\ using phase spaces, symplectic forms, and constraint surfaces. More precisely, we will establish a classical description of an \emph{effective} theory which can be interpreted as the result of integrating out 1-loop quantum fluctuations in a gravitational path integral. Such fluctuations generate a classical anomaly in the effective theory, parametrized by a central charge $c$. Alternatively, this classical anomaly can be understood as a counterterm to cancel the anomaly in the quantum theory, as in the Green-Schwarz mechanism~\cite{GreenSchwarz, Green_Schwarz_Witten_2012} (we give a little more detail on this below, but will explicitly show how it works in~\cite{QuantumAreaTime}).

\end{itemize}

A key technical point is that in the classical theory, the phase space of the dressing time can be identified with the cotangent bundle of the diffeomorphism group, equipped with the corresponding Kirillov-Kostant-Souriau (KKS) symplectic form~\cite{Guillemin:1990ew,Woodhouse:1980pa}. One of the main results of this paper is that the anomaly in the effective theory can be accounted for by deforming the classical phase space to the cotangent bundle of the Virasoro algebra, with its associated KKS form~\cite{Alekseev:2022qrq}. 
This leads to the existence of additional gauge-invariant degrees of freedom in the effective theory relative to the tree-level one~\cite{Henneaux2020}; the phase space of these additional degrees of freedom is a coadjoint orbit of the Virasoro group~\cite{Alekseev:1988ce,Witten:1987ty,Cotler:2018zff,nair2024notecoherentstatesvirasoro,Stanford:2017thb}. The dressing time provides a natural set of coordinates on this coadjoint orbit.

This paper lays the groundwork for the quantization of the dressing time reference frame, where the dressing time is promoted to a genuine QRF~\cite{QuantumAreaTime}. A typical choice of Hilbert space for quantum mechanical QRFs is the space $L^2(G)$ of square-integrable wavefunctions over the relevant symmetry group $G$. In the present case, $G$ is a diffeomorphism group, which does not have square-integrable functions in the usual sense, so the quantization must be done differently. Indeed, the dressing time is a field, and as such should be quantized with a Fock space; thus, it is a kind of K\"ahler QRF~\cite{KahlerQRF}. 

In the effective classical theory, the central charges of the three diffeomorphism representations are given by
\begin{equation}
  \begin{gathered}
    c_{\text{reparametrization}} = c,
    \qquad
    c_{\text{reorientation}} = -c,
    \qquad 
    c_{\text{dressed reparametrization}} = -c,
  \end{gathered}
\end{equation}
It turns out that the quantization of the null ray shifts each of these three charges by different amounts; by picking $c$ appropriately, one can cancel the anomaly for the dressed reparametrization algebra, thus eliminating the extra degrees of freedom in the quantum theory~\cite{QuantumAreaTime}. The resulting structures bear a strong resemblance with the bosonic string, and reparametrization modes in 2D CFTs~\cite{Cotler:2018zff,Alekseev:1988ce,Haehl:2019eae,Nguyen:2021jja,maldacena2016conformalsymmetrybreakingdimensional,Mertens:2017mtv}.

The paper proceeds as follows. In Section~\ref{Section: classical gravity}, we shall describe in some detail the structure of the tree-level theory, including the phase space on a gravitational null ray, and the non-anomalous Raychaudhuri constraint within the decoupling regime in which the constraint splits $C=H_R+H_S$.  We also demonstrate how one may reduce to the phase space of a finite segment of the null ray, and how this leads to the introduction of edge modes. Then we explain how to impose the gauge constraint, and describe in more detail the three diffeomorphism actions: reparametrizations, reorientations, and dressed reparametrizations. We describe the effective corrections to the tree-level structure in Section~\ref{Section: effective}, explaining how the constraint and phase space are modified by anomalous contributions. By extending the tree-level methods of the previous section, we derive the segment phase space to reveal how the anomaly modifies the edge modes. We then explain how reparametrizations, reorientations, and dressed reparametrizations are modified in the effective theory and derive their corresponding central charges, before commenting on the consequences of imposing gauge-invariance in the effective theory. Finally, we conclude the paper in Section~\ref{Section: Conclusion} with a summary of its achievements and a discussion of some open questions.

\section{Tree-level null ray segments}
\label{Section: classical gravity}

In order to establish a solid foundation for the effective anomalous theory, we will first describe in some detail the structure of the purely classical `tree-level' theory, with no anomaly. We start by describing the Raychaudhuri constraint $C$ in a regime in which the spin 0 and radiative fields along a null ray are decoupled, the kinematical phase space of these degrees of freedom, and the gauge transformations generated by $C$. We explain how one may reduce this description to a segment of the null ray by using embedding fields, and show that this leads to the inclusion of edge modes transforming under a corner symmetry group. We introduce the dressing time, and demonstrate its use in constructing dressed observables such as the dressed area element. We derive the Poisson algebra of these dressed observables using both reduced phase space and Dirac bracket techniques, explicitly describing how these give isomorphic results via a so-called dressing map. Finally, we explain the key roles played by three diffeomorphism actions (reparametrizations, reorientations, and dressed reparametrizations).

\subsection{Raychaudhuri constraint in a decoupling regime}
\label{Subsection: decoupling regime}

\begin{figure}
  \centering
  \begin{tikzpicture}[scale=0.9]

    \fill[blue!20,opacity=0.4] (2.5,0) -- (5,7) arc (0:180:5 and 1) -- (-2.5,0) arc (180:0:2.5 and 0.8);
    \foreach \i in {0,1,...,11} {
      \draw[opacity=0.2,dashed] (0,7)++(10+15*\i:5 and 1) -- (10+15*\i:2.5 and 0.8);
    }
    \draw[blue!50,thick] (2.5,0) arc (0:180:2.5 and 0.8);
    \draw[green!50!black,opacity=0.8,thick] (3.5,2.8) arc (0:180:3.5 and 0.88);
    \fill[blue!30,opacity=0.75] (2.5,0) -- (5,7) arc (0:-180:5 and 1) -- (-2.5,0) arc (-180:0:2.5 and 0.8);

    \foreach \i in {12,13,...,23} {
      \draw[opacity=0.3,dashed] (0,7)++(10+15*\i:5 and 1) -- (10+15*\i:2.5 and 0.8);
    }

    \draw[blue!50,thick] (0,7) ellipse (5 and 1);
    \draw[blue!50,thick] (2.5,0) arc (0:-180:2.5 and 0.8);
    \draw[blue!50,thick] (2.5,0) -- (5,7);
    \draw[blue!50,thick] (-2.5,0) -- (-5,7);

    \draw[green!50!black,opacity=0.8,thick] (3.5,2.8) arc (0:-180:3.5 and 0.88);
    \node[green!50!black] at (-2.4,2.6) {\Large$\mathcal{C}$};

    \path (0,7)++(295:5 and 1) -- coordinate[pos=0.9] (a) (295:2.5 and 0.8);
    \path (0,7)++(295:5 and 1) -- coordinate[pos=0.8] (a') (295:2.5 and 0.8);
    \path (0,7)++(295:5 and 1) -- coordinate[pos=0.15] (b) (295:2.5 and 0.8);
    \path (0,7)++(295:5 and 1) -- coordinate[pos=0.05] (b') (295:2.5 and 0.8);
    \draw[very thick,red] (a) -- (b);

    \node[left] at (a) {$v_0$};
    \node[left] at (b) {$v_1$};

    \draw[very thick,-latex] (a) -- (a') node[midway,right] {$\eta_0$};
    \draw[very thick,-latex] (b) -- (b') node[midway,right] {$\eta_1$};
    \fill[red] (a) circle (0.08);
    \fill[red] (b) circle (0.08);

    \node at (0,2.8) {$\Omega,\beta,\varphi,h_{ab}$};
    \node[blue!60!black] at (-2,4.7) {\Large$\mathcal{N}$};

    \node[red] at (2,3) {\Large$\mathcal{I}$};
  \end{tikzpicture}
  \caption{In this paper, we construct an effective classical description of the null ray segments $\mathcal{I}$ of a gravitational caustic-free null surface $\mathcal{N}$. The degrees of freedom are those of matter $\varphi$, perturbative spin 2 gravitational radiation $h_{ab}$, the area element $\Omega$ and its conjugate momentum $\beta$ on cuts $\mathcal{C}$ of the null surface, as well as edge modes: the locations of the endpoints $v_0,v_1$ of the segment, and boost frames $\eta_0,\eta_1$ at each endpoint. }
  \label{Figure: null ray}
\end{figure}
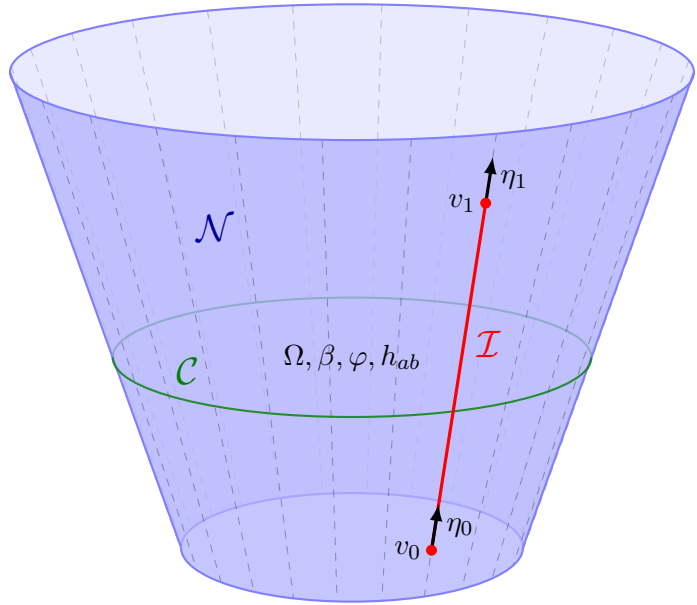

Consider the classical degrees of freedom associated with a null surface $\mathcal{N}$ in gravity. One of the key equations governing these degrees of freedom is the \emph{Raychaudhuri constraint}, which we write in the form
\begin{equation}
  C = \partial_v^2\Omega - \mu \partial_v\Omega + \Omega(\sigma\indices{_a^b}\sigma\indices{_b^a} + 8\pi \GN T^{\text{mat}}_{vv}) = 0.
\end{equation}
Let us explain the terms that appear here. The geometric data intrinsic to $\mathcal{N}$ are its degenerate metric $q_{ab}$ (induced from the spacetime metric) and a nowhere-vanishing null vector field $\ell^a$ (so it satisfies $\ell^aq_{ab}=0$); these comprise its `Carollian structure'~\cite{ciambelli2025foundationscarrolliangeometry}. We consider a partially gauge-fixed phase space with $\delta \ell^a=0$ (referred to in~\cite{Ciambelli:2023mir,Ciambelli_2024} as a `primed' phase space)\footnote{This gauge-fixing amounts to a decoupling of the spin 1 gravitational degrees of freedom. It is only possible here because we are choosing to ignore diffeomorphisms transverse to $\ell$ and their corresponding gauge constraints, including the Damour equation. A full treatment should more carefully account for those constraints, but we leave this for future work.} and define a coordinate $v$ by $\ell=\partial_v$. On constant $v$ cuts $\mathcal{C}$ of $\mathcal{N}$, $q_{ab}$ restricts to a non-degenerate metric which we can use to lower and raise indices; $\Omega$ is the associated infinitesimal area element. The expansion $\theta$ and shear $\sigma_{ab}$ are defined by $\frac12\partial_vq_{ab}=\tfrac12\theta q_{ab}+\sigma_{ab}$ with the condition that the shear is traceless. The quantity $\mu=\kappa-\tfrac12\theta$, where $\kappa$ is the `inaffinity' of $\ell$ (so $\nabla_\ell\ell^b = \kappa \ell^b$), is sometimes called `surface tension'. $T^{\text{mat}}_{vv}$ is the null-null component of the matter stress tensor.

In the following we will consider a linearized regime for the spin 2 part of the gravitational field, meaning the shear $\sigma\indices{_a^b}$. In particular, we assume that $\varkappa h_{ab}$ is very small in the decomposition
\begin{equation}
  q_{ab} = \Omega \qty(q_{ab}^{(0)} + \varkappa h_{ab}),
  \label{Equation: linearised spin 2}
\end{equation}
where $\varkappa = \sqrt{32\pi\GN}$, and $q_{ab}^{(0)}$ is some fixed background degenerate metric with $\delta q_{ab}^{(0)}=\partial_v q_{ab}^{(0)}=\ell^a q_{ab}^{(0)}=0$, and $\det q^{(0)}=1$. One then has
\begin{equation}
  \sigma\indices{_a^b} = \frac12\varkappa \partial_v h\indices{_a^b} + \order{\varkappa ^2},
\end{equation}
where $h\indices{_a^b}=q^{(0)cb}h_{ac}$.
Substituting this into the Raychaudhuri constraint and truncating to quadratic order in $\varkappa h$ yields
\begin{equation}
  C = \partial_v^2\Omega - \mu\partial_v \Omega + 8\pi \GN \Omega \qty(\partial_v h\indices{_a^b}\partial_v h\indices{_b^a} + T^{\text{mat}}_{vv}) = 0.
\end{equation}
At this order, one can treat $h\indices{_a^b}$ as traceless and decompose it into its two polarizations $h_\times,h_+$, so that
\begin{equation}
  \partial_v h\indices{_a^b}\partial_v h\indices{_b^a} = (\partial_v h_\times)^2 + (\partial_v h_+)^2.
\end{equation}
To keep things simple, we will take the matter to just consist of a collection of massless scalar fields $\phi$. It is convenient to absorb the spin 2 polarizations $h_\times,h_+$ into this collection, so that
\begin{equation}
  \partial_v h\indices{_a^b}\partial_v h\indices{_b^a} + T^{\text{mat}}_{vv} = \sum_i (\partial_v\phi_i)^2,
\end{equation}
where the index $i$ runs over the spin 2 and matter contributions.

Let us emphasize that the spin 0 part of the gravitational field $\Omega$ can have arbitrarily large fluctuations in this regime. In the future we hope to understand what happens when one turns back on the higher order terms in $\varkappa h$, and includes other types of matter, perhaps by establishing a perturbation theory around the results of this paper. As we shall see, the present setup is simple enough to be tractable, while simultaneously having sufficient structure to be quite physically and mathematically rich.

In $C$, the scalar fields $\phi_i$ appear to interact with the area element $\Omega$ via a cubic term. It is convenient to eliminate this by defining
\begin{equation}
  \varphi_i := \sqrt{\Omega}\phi_i, \qquad \beta := \mu +4\pi \GN\frac{1}{\sqrt{\Omega}}\partial_v\qty\Big(\sqrt{\Omega}\sum_i\phi_i^2),
  \label{Equation: half-densitisation}
\end{equation}
in terms of which the Raychaudhuri constraint reads
\begin{equation}\label{RC1}
  C = \partial_v^2\Omega - \beta \partial_v\Omega + 8\pi G_N\sum_i(\partial_i\varphi_i)^2.
\end{equation}
We will hereafter refer to the `half-densitized' fields $\varphi_i$ as the radiative degrees of freedom. Note that we have eliminated any interaction terms between the spin 0 degrees of freedom and the radiative ones; indeed, we can write the constraint as $C=H_R+H_S$, where
\begin{equation}
  H_R=\partial_v^2\Omega-\beta\partial_v\Omega
\end{equation}
is the contribution of the spin 0 `reference frame', and
\begin{equation}
  H_S = 8\pi G_N\sum_i(\partial_i\varphi_i)^2
\end{equation}
is the contribution of the radiative `system'.

\subsection{Kinematical phase space}
\label{Subsection: kinematical phase space}

The null surface $\mathcal{N}$ is foliated by its null rays. Since the null rays are out of causal contact with one another, they are dynamically decoupled -- this is the principle of \emph{ultralocality}. It allows each null ray to be treated as an independent physical system, making it possible to perform procedures such as quantization and constraint imposition ray-by-ray. We will exploit this to focus, for the remainder of the paper, on a single null ray $\Ray$.\footnote{Of course, at some later stage the rays must be glued back together to recover the full system. This procedure is likely to involve some uniquely quantum gravitational techniques and phenomena, such as `embadons'~\cite{Ciambelli_2024}, discretization of geometry~\cite{Wieland:2024dop,Wieland:2025qgx}, or fuzzy geometry \cite{Donnelly:2020xgu,Donnelly:2022kfs}}

The `canonical' symplectic 2-form of a gravitational null surface in the `primed' phase space~\cite{Ciambelli:2023mir,Ciambelli_2024} is given by
\begin{equation}
  \bm\Omega^{\text{can}} = \int_{\mathcal{N}}\varepsilon_{\mathcal{N}}^{(0)}\qty(\frac1{8\pi\GN}\qty( \delta\Omega\wedge\delta\mu + \delta\qty(\tfrac12\Omega\sigma^{ab})\wedge\delta q_{ab}) + \omega^{\text{mat}}),
\end{equation}
The volume form used above is $\varepsilon_{\mathcal{N}}^{(0)} = \dd{v} \wedge \varepsilon_{\mathcal{C}}^{(0)}$, where $\varepsilon_{\mathcal{C}} = \Omega \varepsilon_{\mathcal{C}}^{(0)}$ is the volume form associated with $q_{ab}^{(0)}$ on the cuts $\mathcal{C}$. The $\dd{v}$ part gives an integral along each null ray of $\mathcal{N}$, while $\varepsilon_{\mathcal{C}}^{(0)}$ integrates over all null rays. Since we are looking at a single null ray $\Ray$, we will discard this latter integration. The term $\varepsilon_{\mathcal{N}}^{(0)}\omega^{\text{mat}}$ is the symplectic current of the matter fields; in the case at hand $\omega^{\text{mat}}$ is a sum over $\delta(\Omega\partial_v\phi_i)\wedge\delta\phi_i$. This sum by itself does not include the spin 2 modes $h_\times,h_+$, but using the linearized prescription~\eqref{Equation: linearised spin 2} leads to
\begin{equation}
  \frac1{8\pi\GN}\delta\qty(\tfrac12\Omega\sigma^{ab})\wedge\delta q_{ab} = \delta(\Omega\partial_v h_\times)\wedge\delta h_\times + \delta(\Omega\partial_v h_+)\wedge\delta h_+,
\end{equation}
so the symplectic form may be written
\begin{equation}
  \bm\Omega = \int_{\Ray}\dd{v}\qty\Big(\frac1{8\pi\GN} \delta\Omega\wedge\delta\mu +  \sum_i \delta(\Omega \partial_v\phi_i)\wedge\delta\phi_i),
  \label{Equation: symplectic form without edge modes bad}
\end{equation}
where the $\sum_i$ includes the radiative degrees of freedom and the matter fields. Making the change of variables $\mu \to \beta$ and $\phi_i\to\varphi_i$ defined in~\eqref{Equation: half-densitisation}, one finds that the symplectic structure decouples
\begin{equation}
  \bm\Omega = \int_{\Ray}\dd{v}\qty\Big(\frac1{8\pi\GN} \delta\Omega\wedge\delta\beta + \sum_i \partial_v(\delta\varphi_i)\wedge\delta\varphi_i),
  \label{Equation: symplectic form without edge modes}
\end{equation}
up to a boundary term which we can safely ignore by imposing the falloff conditions
\begin{equation}
  \varphi_i=o(1/v),\qquad \Omega=\order{v},\qquad \beta=o(1/v^2) \qq{as} v\to\pm\infty.
  \label{Equation: field falloffs}
\end{equation}

The advantage of writing the symplectic form as~\eqref{Equation: symplectic form without edge modes} should be clear: the spin 0 fields $\Omega,\beta$ are explicitly symplectically decoupled from the radiative fields $\varphi_i$, so the Poisson brackets between them vanish, and importantly the field $\beta$ commutes with itself:
\begin{equation}
  \pb{\Omega(v)}{\varphi_i(v')}=\pb{\beta(v)}{\varphi_i(v')}=0, \qquad \pb{\beta(v)}{\beta(v')}=0.
\end{equation}
This structure is only valid after the change of variables to $\beta$ and the half-densitized radiative fields $\varphi_i$. Indeed, the original radiative fields $\phi_i$ do not commute with $\mu$, because they contribute to~\eqref{Equation: symplectic form without edge modes bad} with a term proportional to $\Omega$. Moreover, $\mu$ doesn't commute with itself~\cite{Ciambelli:2023mir}:
\begin{equation}
  \pb{\mu(v)}{\phi_i(v')}\neq0, \qquad \pb{\mu(v)}{\mu(v')}\neq0.
\end{equation}
In this way, the change of variable $(\mu,\phi_i) \to(\beta,\varphi_i)$  identifies a spin $0$ sector which commutes with the matter and radiative sectors.
Thus, the decomposition $C=H_R+H_S$ amounts to a decoupling consistent with the phase space structure of the theory, because $H_R$ involves only $(\Omega,\beta)$, and $H_S$ involves only $\varphi_i$; thus $\pb{H_R}{H_S}=0$. This dramatically simplifies the solution of the constraint, since the spin $0$ fluctuations don't back-react on the radiative degrees of freedom; the constraint $C=H_R +H_S=0$ only creates correlations (and, in the quantum theory, entanglement) between the spin $0$ and the radiative data.
\subsection{Gauge transformations}
\label{Subsection: gauge transformations}

Let us define the \emph{stress tensor}
\begin{equation}\label{RC2}
  T:= \frac{C}{8\pi\GN},
\end{equation}
where $C$ is defined in \eqref{RC1}. Suppose $f$ is a function satisfying
\begin{equation}
  f = o(1),\qquad \partial_vf = o(1/v), \qq{as} v\to\pm\infty.
  \label{Equation: diffeo falloffs}
\end{equation}
It is straightforward to check using the symplectic form that $T_f:=\int_{\Ray}\dd{v}fT$ is then the generator of the following transformation
\begin{equation}\label{varOb}
  \delta_f \Omega = f\pa_v  \Omega, \qquad \delta_f\beta = \partial_v(f\beta) +\pa_v^2f, \qquad \delta_f \varphi_i = f\partial_v\varphi_i.
\end{equation}
Indeed, we have
\be
I_{\delta_f}\Omega = - \delta T_f.
\ee

Let $F=\exp(f\partial_v)$ denote the (orientation-preserving) diffeomorphism generated by the vector field $f\partial_v$. Then~\eqref{varOb} may be integrated to the finite transformation
\begin{equation}\label{Equation: F action}
  F\triangleright\Omega = \Omega\circ F, \qquad
  F\triangleright \beta = (\partial_v F) \beta \circ F + \frac{\partial_v^2 F}{\partial_v F}, \qquad F\triangleright \varphi_i = \varphi_i\circ F,
\end{equation}
In~\cite{Ciambelli:2023mir,Ciambelli_2024}, this was called a `primed diffeomorphism'. We will just refer to it as a diffeomorphism or reparametrization. It is the combination of an ordinary diffeomorphism tangent to $\ell$ with a rescaling of $\ell$ to preserve the condition $\delta \ell=0$. These are gauge transformations in gravity. Note that $\beta$ transforms like the $v$ component of a connection 1-form (the second term above comes from the rescaling of $\ell$ in the inaffinity $\kappa$).

\subsection{Reducing to a segment with edge modes}\label{edgem}

\begin{figure}
  \centering
  \begin{tikzpicture}
    \draw[decorate,decoration={brace,raise=6pt}] (0.675,1.125) -- (2.025,3.325) node[midway,left=6pt,shift={(0,8pt)}] {\Large $I$};
    \begin{scope}[yellow!40!black]
      \draw[line width=0.9pt,densely dotted,-{Stealth[length=1.5mm]}] (0,0) -- (3,5);
      \fill (0.6,1) circle (0.05) node[below right=-1pt] {\small$0$};
      \fill (2.1,3.5) circle (0.05) node[below right=-1pt] {\small$1$};
    \end{scope}
    \node[above] at (3,5) {$\mv $};
    \node[above left=-1pt,gray] at (0.24,0.4) {\footnotesize$\bar I$};
    \node[above left=-1pt,gray] at (2.4,4) {\footnotesize$\bar I$};

    \begin{scope}[rotate=-15]
      \begin{scope}[shift={(3,0.5)}]
        \path (3.6,6) --+ (105:-0.1) coordinate (a);
        \path (3.6,6) --+ (15:1) coordinate (c);
        \path (c) --+ (105:-0.1) coordinate (b);
        \path[shading=axis, left color=blue!25, right color=blue!0, shading angle=35] (15:6) -- (0,0) -- (3.6,6) .. controls (a) and (b) .. (c)  -- cycle;
        \draw[decorate,decoration={brace,raise=6pt}] (2.325,3.825) -- (0.675,1.125) node[midway,right=6pt,shift={(0,-8pt)}] {\Large $\mathcal I$};
        \begin{scope}[red!60!black]
          \draw[dashed,-{Stealth[length=2mm]},gray!80!blue,line width=1pt] (0,0) -- (3.6,6);
          \draw[line width=1.5pt,red] (0.6,1) -- (2.4,4);
          \fill (0.6,1) circle (0.08) node[above left=-1pt] {\small$v_0$};
          \fill (2.4,4) circle (0.08) node[above left=-1pt] {\small$v_1$};
        \end{scope}
        \node[above] at (3.6,6) {$v$};
        \node[below right=-1pt,gray] at (0.36,0.6) {\footnotesize$\bar {\mathcal I}$};
        \node[below right=-1pt,gray] at (3,5) {\footnotesize$\bar {\mathcal I}$};
      \end{scope}
    \end{scope}

    \draw[gray,thick,double distance=3.5pt] (1.4,2.1) .. controls (2.3,1.5) and (3.7,2.5) .. (4.85,1.6) node[near end,above=2pt,shift={(-3pt,0)},black] {$X$}  node[near start,below=2pt,shift={(-3pt,0)},black] {$V$};
    \draw[{Latex[length=2mm]}-,gray,shift={(-2.5pt,-1pt)}] (1.4,2.1) -- (1.42,2.088);
    \draw[{Latex[length=2mm]}-,gray,shift={(2.5pt,0.8pt)}] (4.85,1.6) -- (4.83,1.614);

    \begin{scope}[shift={(1.2,2)}]
      \draw[double distance=3.5pt,gray,thick] (1.3,2) .. controls (2,1.5) and (4.65,3.1) .. (5.95,1.9) node[near end,above=1pt,black] {$\bar X$} node[near start,below=2pt,shift={(-3pt,0)},black] {$\bar V$};
      \draw[{Latex[length=2mm]}-,gray,shift={(-2.5pt,-1pt)}] (1.3,2) -- (1.32,1.988);
      \draw[{Latex[length=2mm]}-,gray,shift={(2.5pt,0.8pt)}] (5.95,1.9) -- (5.93,1.914);
    \end{scope}
  \end{tikzpicture}
  \caption{We use embedding fields $X:I\to \mathcal{I}$ and $\bar{X}:\bar I\to \bar{\mathcal{I}}$ to construct the phase space of the null ray segment $\mathcal{I}$. Here, $I=[0,1]\in\RR$ is a reference interval for $\mathcal{I}$ via $X$, while the complement $\bar I$ is a reference for the complement $\bar{\mathcal{I}}$ via $\bar X$. On $I\cup\bar I$ we use the coordinate $\mv$, while on $\mathcal{I}\cup\bar{\mathcal{I}}$ we use the coordinate $v$. We will take $X$ to be the inverse of the `dressing time' $V:\mathcal{I}\to I$, but leave $\bar X$ general, only requiring $X(a)=\bar X(a)=v_a$ at $a=0,1\in\partial I$. The mismatches between the derivatives of $X$ and $\bar X$ at $\partial I$ manifest as edge modes in the segment phase space. }
  \label{Figure: embedding}
\end{figure}
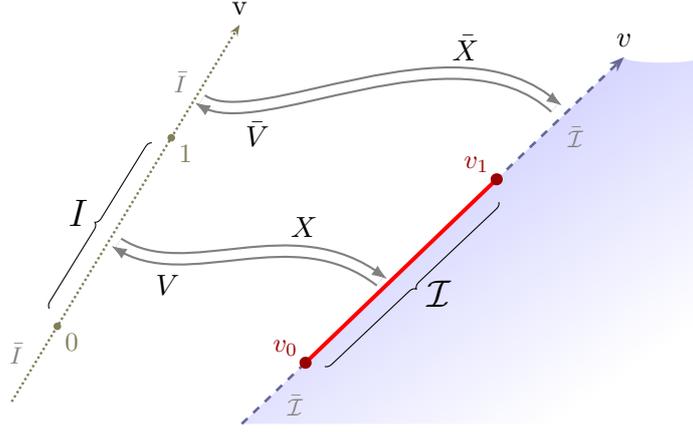

So far we have described the phase space of the full null ray $\Ray$. From this we will now derive the phase space of a subregion of the ray, i.e.\ a \emph{segment} $\Segment\subset\Ray$. Let us use $v=v_a$, $a=0,1$ to denote the endpoints of $\Segment$, with $v_0<v_1$.

The basic assumption we shall make in the following is that observables $\mathcal{O}$ supported in $\Segment$ commute with observables $\bar{\mathcal{O}}$ supported in its complement $\SegmentBar=\Ray\setminus\Segment$, i.e.\ their Poisson bracket vanishes, $\pb*{\mathcal{O}}{\bar{\mathcal{O}}}=0$.

In infinite dimensions, even when the symplectic structure is non-degenerate, the Poisson bracket is only defined for certain observables\footnote{In other words, even if $\Omega$ is non-degenerate, the map $x\mapsto [y\mapsto \Omega(x,y)]$ is injective but not surjective. This means that there could be phase space observables $O$ such that $\delta \mathcal{O}$ is not of the form $\Omega(\cdot, y)$ for any Hamiltonian vector field $Y$~\cite{Chandrasekaran:2023vzb}.}~\cite{Prabhu:2022zcr}.
The bracket exists when there is a Hamiltonian vector field $\delta_{\mathcal{O}}$, meaning $I_{\delta_{\mathcal{O}}}\bm\Omega =-\delta \mathcal{O}$ (here $I$ denotes phase space contraction).
The Poisson bracket of two such observables can be found by contracting their corresponding vector fields into the symplectic form:
\begin{equation}
  \pb{\mathcal{O}_1}{\mathcal{O}_2} = \mathbf{\Omega}(\delta_{\mathcal{O}_1},\delta_{\mathcal{O}_2}) = \delta_{\mathcal{O}_1}\mathcal{O}_2.
  \label{Equation: symplectic Poisson relation}
\end{equation}
The convention chosen here means that $O\to \delta_{\mathcal{O}}$ is an algebraic morphism, i.e. $\delta_{\pb{\mathcal{O}_1}{\mathcal{O}_2}} =[\delta_{\mathcal{O}_1},\delta_{\mathcal{O}_2}]$.\footnote{A simple proof of this is
  \be
  I_{\delta_{\{\mathcal{O}_1, \mathcal{O}_2\}}}\bm\Omega =
  -\delta \{O_1,O_2\}= -\delta L_{\delta_{\mathcal{O}_1}}\mathcal{O}_2=
  -L_{\delta_{\mathcal{O}_1}} \delta \mathcal{O}_2
  = L_{\delta_{\mathcal{O}_1}} I_{\delta_{\mathcal{O}_2}} \bm\Omega =
  I_{[\delta_{\mathcal{O}_1},\delta_{\mathcal{O}_2}]}\bm\Omega,
  \ee
where we have used that $L_{\delta_{O_{1,2}}}\bm\Omega=0$.}

In theories with gauge symmetry, observables $\mathcal{O}$ must commute with the constraints (so here $\pb{\mathcal{O}}{T_f}=0$); equivalently they must be invariant under gauge transformations (so $\delta_f\mathcal{O}=0$). Suppose we wish to compute the Poisson bracket of two observables $\mathcal{O}_1,\mathcal{O}_2$ supported in $\Segment\subset \Ray$, with associated Hamiltonian vector fields $\delta_{\mathcal{O}_1}, \delta_{\mathcal{O}_2}$.
By the assumption that observables in $\Segment$ commute with those in $\SegmentBar$, the field variations $\delta_{\mathcal{O}_1},\delta_{\mathcal{O}_2}$ must leave invariant any gauge invariant observable supported in $\SegmentBar$, which implies that $\delta_{\mathcal{O}_1},\delta_{\mathcal{O}_2}$ must be indistinguishable from gauge transformations (here primed diffeomorphisms) in $\SegmentBar$~\cite{Pulakkat:2025eid}. To compute the Poisson bracket of observables supported in $\Segment$ with~\eqref{Equation: symplectic Poisson relation}, it therefore suffices to restrict the symplectic form to field variations with this property. Since the Poisson bracket should ultimately be evaluated on-shell of the constraints, we are free to also impose $T\onsheq 0$ in $\SegmentBar$ (where $\onsheq$ denotes on-shell equality). The result is a symplectic form suitable for use on the segment $\Segment$.

Let us write the symplectic form for $\Ray$ as $\bm\Omega=\delta\Theta$, where the symplectic potential is
\begin{equation}
  \Theta = \int_{\mathbb{R}} \dd{v}\qty\Big(\tfrac1{8\pi \GN}\Omega \delta \beta + \sum_i \partial_v\varphi_i\delta\varphi_i).
\end{equation}
We decompose this as ${\Theta}=\hat{\Theta}_{\Segment} + {\Theta}_+ - {\Theta}_-$ where
\begin{align}
  \hat{\Theta}_{\Segment} &= \int_{\Segment}\dd{v}\qty\Big(\tfrac1{8\pi \GN}\Omega \delta \beta + \sum_i \partial_v\varphi_i\delta\varphi_i),\\
  {\Theta}_a &= \int_{v_a}^{\epsilon_a\infty}\dd{v}\qty\Big(\tfrac1{8\pi \GN}\Omega \delta \beta + \sum_i \partial_v\varphi_i\delta\varphi_i),
\end{align}
and $\epsilon_0=-1$, $\epsilon_1=+1$.

Let us first focus on the terms $\Theta_a$. We assume that the field variations outside $\Segment$ are equal to some gauge transformation,\footnote{As we will explain in Section~\ref{externalD}, this is only strictly true for $\delta\beta$ and $\delta\varphi_i$. One must allow more general variations for $\Omega$ that include edge mode variations.} so
\begin{equation}
  \delta\beta = \partial_v(f\beta) + \partial_v^2 f, \qquad \delta\varphi_i = f\partial_v \varphi_i
\end{equation}
for some $f$ satisfying the falloffs~\eqref{Equation: diffeo falloffs}. Substituting this into $\Theta_a$ and integrating by parts, one finds
\begin{equation}
  \Theta_{a}=   \tfrac1{8\pi\GN}\left[\pa_v\Omega f- \Omega (\pa_vf+\beta f)  \right](v_a) + \int_{v_{a}}^{\epsilon_a\infty} f T \rd v.
  \label{Equation: Theta a}
\end{equation}
On-shell of the constraint $T=0$ outside $\Segment$, each $\Theta_a$ therefore simplifies to a boundary term at $v_a$.

In order to account for these boundary variations in the symplectic structure for $\Segment$, one needs to introduce edge modes degrees of freedom $(v_a,\bar\eta_{a})$ which transform under diffeomorphisms as
\be \label{varvet}
\delta_f v_a = - f(v_a), \qquad
\delta_f \bar{\eta}_a =  \pa_v f (v_a).
\ee
Here, $v_a$ is the same $v_a$ defining the region $\mathcal{I}$. In the language of \cite{Donnelly:2016auv}, this corresponds to introducing an embedding field $\bar{X}: \bar{I} \to \bar{\mathcal{I}}$ which maps the complement\footnote{$\bar{I}= \mathbb{R}\backslash I$ and similarly for $\bar{\mathcal{I}}$.} of a reference interval $I=[0,1]$ onto the complement of $\mathcal{I}=[v_0,v_1]$, and defining
\be
v_a = \bar{X}(a), \qquad \bar{\eta}_{a} = -\ln\qty\big(
\pa_{\mathrm{v}} \bar{X})(a).
\ee
Here and henceforth we use $\mv$ to denote the coordinate on the reference space containing $I$ (in contrast to $v$, which we use as the coordinate on the ray $\Ray$). Under a gauge transformation, the embedding field transforms according to a diffeomorphism acting on the  target of  $\bar{X}$: $\delta_f \bar{X} = -f\circ \bar{X}$.\footnote{Hence, for example, $\delta_f (\pa_{\mathrm{v}}\bar{X})=-
\pa_{\mathrm{v}}(f\circ \bar{X}) =- \pa_{\mathrm{v}}\bar{X}(\pa_v f) \circ \bar{X}$.}

A finite gauge transformation acts as $F\triangleright \bar X = F^{-1}\circ \bar X$, so
\begin{equation}
  F\triangleright v_a = F^{-1}(v_a), \qquad F\triangleright \bar\eta_a = -\ln\partial_v F^{-1}(v_a).
  \label{Equation: edge mode finite diffeo}
\end{equation}
The first of these implies $F\triangleright\Segment = F^{-1}(\Segment)$. This is sensible, because under a primed diffeomorphism $F$ acting on the fields as in~\eqref{Equation: F action} the support of an observable $\mathcal{O}$ changes via $\operatorname{supp}(\mathcal{O})\mapsto F^{-1}(\operatorname{supp}(\mathcal{O}))$.\footnote{Here $\operatorname{supp}(\mathcal{O})$ is shorthand for $\operatorname{supp}(\fdv{\mathcal{O}}{\Phi})$, where $\fdv{\mathcal{O}}{\Phi}$ is the functional derivative of $\mathcal{O}$ with respect to the fields (collectively denoted $\Phi$). Note that $\operatorname{supp}(\mathcal{O})$ depends on the field configuration and hence can vary over phase space.} To have a notion of observables supported in $\Segment$ that is consistent with covariance, $\Segment$ must transform in the same way. This is achieved by the addition of the edge variables $v_a$ to the phase space (called dynamical cuts in~\cite{kirklin2024generalisedsecondlawsemiclassical}).

Therefore, the symplectic structure for gauge-invariant observables supported in $\mathcal{I}$ can be written as
\begin{equation}
  \Theta_{\mathcal{I}} = \int_{\Segment}\dd{v}\qty\Big(\tfrac1{8\pi \GN}\Omega \delta \beta + \sum_i \partial_v\varphi_i\delta\varphi_i) - \sum_{a} \epsilon_a \left(p_a \delta v_a + \omega_a (\delta\bar{\eta}_a -\beta_a \delta v_a)\right).
  \label{Equation: I symplectic potential}
\end{equation}
where $\beta_a=\beta(v_a)$, and the momenta conjugate to the edge mode variables are given by
\be
p_a  =
\frac{\pa_{v_a}\Omega(v_a) }{8\pi\GN} , \qquad \omega_a = \frac{\Omega(v_a)}{8\pi\GN}.
\ee

Recall that primed diffeomorphisms are gauge transformations only if $f,\partial_v f$ fall off at the asymptotic boundaries $v\to\pm\infty$. On the other hand, there is no requirement that $f,\partial_v f$ vanish at the boundaries $v_a$ of $\Segment$. On $\Segment$, at each boundary $v_a$, there is in this way a distinguished residual two-dimensional group of gauge transformations labelled by the values of $f,\partial_v f$. This group is the affine group $\Aff = \mathbb{R}\ltimes\mathbb{R}$, whose elements $(e^{\bar\eta},v)$ correspond to affine transformations $x\mapsto v+xe^{\bar\eta}$. The product and inverse may be written
\be
g'g = (e^{\bar\eta+\bar\eta'}, e^{-\bar\eta} v'  + v), \qquad g^{-1}= (e^{-\bar\eta} ,-e^{\bar\eta} v).
\ee
The edge modes label an element of $\operatorname{Aff}$, and the boundary terms in~\eqref{Equation: I symplectic potential} may be understood as the invariant symplectic forms on the corresponding Lie algebra. Indeed, the Maurer-Cartan form is
\be
\rd g g^{-1}= (\rd \bar\eta,  e^{\bar\eta} \rd v),
\ee
While the group symplectic form is
\be \label{Coadjointorb}
\langle \pi, \rd g g^{-1} \rangle = \tilde{p} e^{\bar\eta} \rd v + \omega \rd \bar\eta,
\ee
where $\pi=(\omega, \tilde{p}) \in \mathfrak{g}^*$ is a dual Lie algebra element.
This matches the boundary term in~\eqref{Equation: I symplectic potential} when $\beta=0$, if we identify $\tilde{p}_a = e^{-\bar\eta_a}p_a = \frac1{8\pi\GN}\frac{\partial_v \Omega}{\partial_v \bar V}(v_a) $ where $\bar V=\bar X^{-1}$.
Note that the edge mode transformation~\eqref{Equation: edge mode finite diffeo} means that the diffeomorphism invariant edge modes are $\omega_a$ and $\tilde{p}_a$. These are exactly the combinations that appear in the Lie algebra symplectic form \eqref{Coadjointorb} as elements of the dual to the affine Lie algebra. This is what one  expects from the general theory~\cite{Donnelly:2016auv}.

\subsection{Gauge-invariance and dressing}
\label{Subsection: dressing}

In order to construct quasi-localized gauge invariant observables with support inside $\mathcal{I}$, let us now introduce the \emph{dressing time} $V$~\cite{Ciambelli_2024}, which is a field internal to the interval. $V \in \operatorname{Diff}^+(\mathcal{I}\to I)$ is an orientation preserving diffeomorphism that maps the segment $\mathcal{I}=[v_0,v_1]$ into the reference interval $I =[0,1]$.

The dressing time is defined from $(\beta,v_a)$ by the differential equation
\be
\frac{\pa_v^2 V}{\pa_v V} = \beta.
\ee
This is a second order differential equation, and accordingly $\beta$ itself only determines $V$ up to the two-parameter group of affine transformations $V\to A V+B$. The edge mode boundary conditions
\be \label{bc}
V(v_a)=a, \qquad a=0,1,
\ee
fix this ambiguity.
$V$ is explicitly given by
\be
V(v)=
\int_{v_0}^v \exp\qty\bigg(\int_{v_0}^{v'} \beta(v'')\dd{v''}) \dd{v'} \Bigg\slash\int_{v_0}^{v_1}\exp\qty\bigg(\int_{v_0}^{v'} \beta(v'')\dd{v''}) \dd{v'}.
\ee
One may readily derive that under a primed diffeomorphism the dressing time transforms as a scalar:
\begin{equation}
  F\triangleright V = V\circ F, \qq{or infinitesimally} \delta_f V = f\partial_v V.
\end{equation}
Note that diffeomorphisms act on the right of the dressing field; in the following we will sometimes refer to this action as the right action. 
The inverse of the dressing time, which we denote $X=V^{-1}$, is an embedding field of the kind described in~\cite{Donnelly:2016auv, Ciambelli:2021nmv, Freidel:2021dxw}.

Using $\delta\beta = \partial_v\left(\frac{\partial_v\delta V}{\partial_v V}\right)$, and integrating by parts, one may rewrite the symplectic potential~\eqref{Equation: I symplectic potential} in the form
\begin{equation}
  \Theta_{\mathcal{I}} = \int_{v_0}^{v_1}\dd{v}\qty\bigg( \tau\frac{\delta V}{\pa_v V} + \sum_i \partial_v\varphi_i\delta\varphi_i) + \Theta_{\pa \mathcal{I}},
  \label{Equation: I symplectic potential 2}
\end{equation}
where
\begin{equation}
  \Theta_{\pa \mathcal{I}} = \frac1{8\pi\GN}\qty[ \Omega \frac{\partial_v\delta V}{\partial_v V}- \frac{\partial_v\Omega}{\partial_v V}\delta V ]_{v_0}^{v_1} - \sum_{a} \epsilon_a \left(p_a \delta v_a +\omega_a (\delta\bar\eta_a -\beta_a \delta v_a)\right).
  \label{Equation: partial I symplectic potential 2}
\end{equation}
Here, we have defined $\tau$ as the spin 0 stress tensor
\begin{equation}
  \tau = \frac{\pa_v V}{8\pi\GN} \partial_v\qty(\frac{\partial_v\Omega}{\partial_v V}) = \frac{1}{8\pi\GN}\qty(\partial_v^2\Omega-\beta\partial_v\Omega).
\end{equation}
This field $\tau$ transforms under primed diffeomorphisms like a rank 2 tensor:
\begin{equation}
  F\triangleright \tau = (\partial_v F)^2\tau\circ F, \qq{or infinitesimally} \delta_f \tau = f\partial_v \tau + 2\pa_vf \tau .
\end{equation}
In terms of $(\tau,V)$, the full stress tensor may be written
\begin{equation}
  T = \tau + \sum_i(\partial_v\varphi_i)^2,
\end{equation}
One observes from~\eqref{Equation: I symplectic potential 2} that $\tau$ plays the role of the variable conjugate to the diffeomorphism Maurer-Cartan form
\be \label{Maurer-cartan}
\theta_{\text{diff}} =\delta X \circ X^{-1} = -\frac{\delta V}{\pa_v V},
\ee
This also means that the 1-form variable
\be
\Pi:= \frac{\tau}{\pa_v V} = \partial_v\qty(\frac{\partial_v\Omega}{\partial_v V}), 
\ee
is the momentum conjugate to $V$.

The Maurer-Cartan form $\theta_{\text{diff}}$ satisfies the zero curvature condition
\begin{equation}
  \delta \theta_{\text{diff}} + \theta_{\text{diff}} \wedge \pa_v \theta_{\text{diff}} = \delta\theta_{\text{diff}} + \frac12[\theta_{\text{diff}},\theta_{\text{diff}}] =0,
  \label{Equation: zero curvature}
\end{equation}
where the bracket is the Witt bracket $[f,g]=f\pa_vg-g\pa_vf$. It is invariant under what we shall call \emph{reorientations} $\delta_g V =  g\circ V  $.
These reorientations act on the left of $V$,\footnote{The left and right actions refer to how they act on the dressing field $V$. Diffeomorphisms $\delta_f V= f \pa_v V $ where $f\in \mathrm{Vect{\mathcal{I}}}$ act on the right while reorientations $\delta_g V = g\circ V$, with $g\in \mathrm{Vect{I}}$ act on the left. } so the Maurer-cartan form $\theta_{\text{diff}}$ is \emph{left-invariant}. This can be checked explicitly as 
\be 
L_{\delta_g}(\delta V)=  (\pa_{\mv}g\circ V) \delta V, \qquad    \qquad \delta_g (\pa_v V)= (\pa_{\mv}g \circ V) \pa_v V.
\ee 
where $L_{\delta_g}$ denotes the field space Lie derivative along the field space vector field $\delta_g$, While $I_{\delta_g}$ denotes the field space interior product.
We shall see below that reorientations of the dressing time frame correspond to physical symmetries rather than gauge transformations.
Note finally that under diffeomorphisms $\theta_{\text{diff}}$ transforms as a vector:
\be 
L_{\delta_f}(\theta_{\text{diff}})=   f \pa_v \theta_{\text{diff}}-\theta_{\text{diff}} \pa_v f
= [f, \theta_{\text{diff}}], \qquad I_{\delta_f}\theta_{\text{diff}} = - f .
\ee 

The boundary term~\eqref{Equation: partial I symplectic potential 2} in the symplectic potential simplifies significantly.
To show this, we need to relate the value of the variation of  $V$ and its derivative at $v_a$ to the edge mode variables.
We have $V(v_a)=a$, and define\footnote{This is equivalent to the edge mode definition $V^{-1}(a)=v_a$ and
  $\eta_a =-
  \ln \pa_{\mathrm{v}} (V^{-1})(a)
  $, where $X=V^{-1}$  plays the role of the embedding field internal to $I$. With this data one can define
$\eta(v)=\eta_0+\int_{v_0}^v\beta$ which transforms as $\delta_f \eta(v) = \pa_vf +\beta f$. This will be useful in Section~\ref{Section: effective}.}
\be
\eta_a =  \ln[\pa_v V(v_a)].
\ee
In this way, $X=V^{-1}$ plays the role of an embedding field internal to $I$.\footnote{While $\bar{X}$ introduced earlier is the embedding field for $\bar{I} = \mathbb{R}\backslash I$.}
From this, we can evaluate the variations
\be
(\delta V)(v_a) =-\pa_v V (v_a) \delta v_a, \qquad \delta\eta_a=
\delta \ln [\pa_v V(v_a)]=\frac{\pa_v\delta V}{\pa_v V}(v_a) + \beta(v_a) \delta v_a,
\ee
from which we have\footnote{A possible generalization is to chose $I=[\mv_0,\mv_1]$ to be different from the complement of $\bar{I}$ given by $[0,1]$ and impose that $V(v_a)=\mv_a$. This still imposes that $X(I)=\mathcal{I}$ is complementary to $\bar{\mathcal{I}}$, but it allows the variation
$\delta V(v_a) = -\pa_v V(v_a) \delta v_a + \delta \mv_a$. This adds a term to \eqref{bdOm} given by $-\tilde{p}_a \delta \mv_a$, where $\tilde{p}_a = p_a e^{-\eta}_a$. For the moment we assume $v_a=a$ and this term is not present.}
\be \label{bdOm}
\frac1{8\pi\GN}\qty[ \Omega \frac{\partial_v\delta V}{\partial_v V}- \frac{\partial_v\Omega}{\partial_v V}\delta V ]_{v_0}^{v_1}
= \sum_a \epsilon_a[ \omega_a(\delta\eta_a - \beta_a \delta v_a) + p_a \delta v_a] .
\ee
Adding the two terms, we see that the terms proportional to $\delta v_a$ cancel, and we are simply left with
\begin{equation}\label{Deta}
  \Theta_{\partial\Segment} = \sum_a \epsilon_a \omega_a\delta q_a \qq{where} q_a = \eta_a - \bar\eta_a.
\end{equation}
Thus, the boundary term amounts to a simple pairing of the boundary areas $\omega_a$ with the relative boosts $q_a$ between the internal and external embedding fields.

The edge mode contribution therefore results from the fact that the internal embedding field $X=V^{-1}$ differs, in general, from the external embedding field $\bar{X}$, and thus that $\eta_a$ can be different from $\bar\eta_a$,
even if both embeddings satisfy $v_a=X(a)=\bar{X}(a)$.\footnote{Relaxing this latter condition would be interesting, but goes beyond what we need in this paper.} 
Note that the external embedding field $\bar{X}$ can be chosen arbitrarily. The only conditions we need are the initial conditions at $v_a$ and the asymptotic conditions $\bar V\sim v$ at $\pm\infty$.  If for example we decide to gauge fix $\bar{X}$  to satisfy the dressing time condition  $\beta|_{\SegmentBar}=\frac{\partial_v^2\bar V}{\partial_v\bar V}$, then the asymptotic conditions on $\bar{X}$ will still typically lead to a mismatch between $\eta_a$ and $\bar\eta_a$.
Note that the fact that edge modes can be understood as a difference between internal and external dressings was first emphasized in the context of Maxwell theory in \cite{Araujo-Regado:2024dpr}.

Taking an exterior derivative of the symplectic potential gives the total symplectic form for $\Segment$:
\begin{equation}
  \bm\Omega_{\Segment} = \delta\qty(\int_{\mathcal{I}}\dd{v}\qty\bigg(\tau \frac{\delta V}{\partial_v V} + \sum_i \partial_v\varphi_i\delta\varphi_i))+ \sum_a \epsilon_a \delta\omega_a \wedge \delta q_a.
  \label{Equation: Omega I}
\end{equation}

\subsubsection{Dressed fields, constraint and phase space}

By pulling back fields through the embedding map $X$, one may construct `dressed fields' such as
\begin{equation}\label{dressed}
  \tilde\varphi_i = X^*\varphi_i = \varphi_i\circ X, \qquad \tilde\tau = X^* \tau = (\partial_\mv X)^2 \tau\circ X.
\end{equation}
These are fields on the reference interval $I=[0,1]$. They are gauge-invariant by construction, because under a gauge transformation one has $X\to F^{-1}\circ X$ and e.g.\ $\varphi_i\to F^*\varphi_i$, so overall
\begin{equation}
  \tilde\varphi_i = \varphi_i\circ X \to (F^{-1}\circ X)_* (\varphi_i\circ F) =  \tilde\varphi_i.
\end{equation}

An important fact is that the dressed version of $\beta$ identically vanishes:
\begin{align}
    \tilde\beta= X\triangleright \beta &= (\partial_\mv X)\beta\circ X + \frac{\partial_{\mv}^2X}{\partial_\mv X} 
    = (\partial_\mv X)\qty(\qty(\frac{\partial_v^2V}{\partial_v V})\circ X + \frac{\partial_{\mv}^2X}{(\partial_\mv X)^2}) = 0
    \label{Equation: tilde beta}
\end{align}
where we used the inversion formula $\frac{\partial_v^2V}{\partial_v V} = - \frac{\partial_{\mv}^2X}{(\partial_\mv X)^2}\circ V$. In this sense, the operation of dressing observables with respect to the dressing time is similar in spirit to imposing the local gauge-fixing condition $\beta=0$. But note that this local condition is not sufficient to completely fix the gauge -- indeed there is a residual affine symmetry as noted above. The frame $V$ contains two additional degrees of freedom: the edge modes $v_0,v_1$. We call the choice of frame $V$ a `frame-fixing', to distinguish it from the `gauge-fixing' $\beta=0$. In a general sense, the difference between frame-fixing and local gauge-fixing is precisely the presence of the edge modes, which play the role of boundary conditions required for the solution of the local gauge-fixing condition (typically a differential equation involving the local fields).\footnote{In general, a gauge-fixing is a local condition on the fields that renders the equation of motion hyperbolic. A frame-fixing is a complete characterization of the representative along the gauge orbit. Gauge-fixing is what is needed to define the inverse propagator and the path integral. Frame-fixing is what is needed to define gauge invariant observables. The difference between the two notions is the presence of the edge modes degrees of freedom.}

One may also dress the Raychaudhuri constraint to obtain
\begin{equation}\label{tildeT}
  \tilde T = (\partial_\mv X)^2 T\circ X = \tilde\tau + \sum_i(\partial_\mv \tilde\varphi_i)^2.
\end{equation}
Note that $\tilde T=0$ and $T=0$ are equivalent conditions.

In the symplectic potential, the change of perspective from $\mathcal{I}$ to $I$ in the spin 0 sector may be encoded by the identities
\be
\int_{\mathcal{I}} \dd{v} \tau \frac{\delta V}{\pa_v V} = \int_{{I}} \dd{\mv} \tilde\tau \delta V\circ V^{-1} = - \int_{{I}} \dd{\mv} \tilde\tau \frac{\delta X}{\pa_{\mv}X}.
\label{Equation: left right maurer cartan}
\ee
Thus, `in the perspective of' the dressing time $V$, $\tilde{\tau}$ is paired with the \emph{right}-invariant Maurer-Cartan form of the $\operatorname{Diff}(\mathbb{R})$ group
\be
\tilde{\theta}_{\text{diff}}= - \frac{\delta X}{\pa_\mv X} = \delta V \circ V^{-1}
\ee
Like the left-invariant $\theta_{\text{diff}}$, this form satisfies the zero curvature condition
\be
\delta \tilde{\theta}_{\text{diff}} + \frac12 [\tilde{\theta}_{\text{diff}},\tilde{\theta}_{\text{diff}}]=0.
\label{Equation: zero curvature tilde}
\ee
It is invariant under the right action of diffeomorphims: $L_{\delta_f}\tilde{\theta}_{\text{diff}}=0$.

\subsubsection{Dressed area element}
\label{Section: Dressed area}

It may be confirmed that the dressed area $\tilde\Omega = \Omega \circ X$ is related to $\tilde\tau$ by
\begin{equation}
  \tilde\tau = \frac1{8\pi\GN}\partial_\mv ^2\tilde\Omega
\end{equation}
The boundary conditions $\frac1{8\pi\GN}\tilde{\Omega}(a)=\omega_a$ fix the integration constants. To give a more explicit expression, let us introduce
\begin{equation}
  \Greens{\mv}{\mv'} = \frac12\qty\Big(\abs{\mv-\mv'} +2 \mv\mv'-\mv-\mv').
  \label{Equation: Green's}
\end{equation}
This Green's function is symmetric under the exchange $\mv\leftrightarrow \mv'$, obeys $\pa_{\mv}^2 \Greens{\mv}{\mv'} = \delta(\mv-\mv')$, and satisfies the boundary condition \be
\Greens{\mv}{1}=\Greens{\mv}{0}=0.
\ee
Using this we can write $\tilde\Omega$ in the segment $I$ as
\begin{align}\label{area}
  \frac{\tilde\Omega_I(\mv)}{8\pi\GN} &= \mv \omega_1+(1-\mv)\omega_0  + \int_{0}^1 \dd{\mv'}\Greens{\mv}{\mv'}\tilde\tau(\mv').
\end{align}
Using  the identity
\be
\int_0^1 (a\mv+ b) \tilde{\tau}(\mv) \rd \mv = (a+b) \tilde{p}_1 - b \tilde{p}_0- b(\omega_1-\omega_0),
\ee
it is direct to check that $\tilde\Omega_I$ satisfies the boundary conditions
\be
\frac{\tilde\Omega_I(a)}{8\pi\GN}= \omega_a, \qquad
\frac{\pa_{\mv}\tilde\Omega_I(a)}{8\pi\GN}= \tilde{p}_a.
\ee

Note that this expression only agrees with the area element $\tilde{\Omega}(\mv)$ when $\mv\in I$. Outside $I$, the area depends through the Raychaudhuri equation on the value of the radiative fields in the exterior $\varphi_i|_{\SegmentBar}=\bg{\varphi}_i \circ \bar{X}$, which we have left unspecified (here $\bg{\varphi}_i$ are some background fields which are not part of the phase space of the segment $\Segment$, see Subsection~\ref{externalD}). Indeed, evaluating $\tilde{\Omega}_I$ for $\mv\in\bar I$ yields
\begin{align}
  \frac{\tilde{\Omega}_I(\mv)}{8\pi\GN}  = \omega_a + \tilde{p}_a (\mv-a), \,\,  \mathrm{when}\quad  \mv \in[a,\epsilon_a \infty[.
\end{align}
Therefore, $\tilde \Omega=\tilde \Omega_I$ on all of $\Ray$ only when there is no radiation on $\bar{\mathcal{I}}$.

\subsection{Algebraic structure}
\label{Subsection: algebraic structure}

Let us now show how to compute the brackets of the gauge-invariant observables constructed above. Generally speaking there are various equivalent ways to do this, and the choice of procedure depends on whether one is inclined to think about things in terms of Poisson structures or symplectic structures, and whether one wants to do gauge-fixing or not.

In terms of Poisson structures, one can just restrict to gauge-invariant observables, and employ the Poisson bracket obtained by inverting the kinematical symplectic form~\eqref{Equation: Omega I}. On the other hand, one can proceed at the symplectic level by constructing a \emph{reduced} phase space. This consists of restricting to the subspace of the kinematical phase space in which the constraints are satisfied, and then quotienting by the action generated by the first-class constraints.

Alternatively, one can impose a gauge-fixing condition in addition to the gauge constraint $C=0$. Here, a very natural gauge-fixing condition is to make the embedding field trivial, $X(\mv )= \mv $ (this is a frame-fixing in the sense described above). All functions of the remaining degrees of freedom are then observables, and the Poisson bracket is replaced by the Dirac bracket.\footnote{One can also do gauge-fixing at the level of the reduced phase space, by simply including the gauge-fixing condition among the set of constraints considered.}

In the following, we will explicitly demonstrate these different approaches, and then show that they give isomorphic results, where the isomorphism is a `dressing map' taking bare observables to their gauge-invariant counterparts.

\subsubsection{Poisson brackets of dressed fields}

In the present case, the full set of gauge-invariant observables in $\Segment$ consists of functions of $\omega_a,q_a$ and the dressed fields $\tilde\tau,\tilde\varphi_i$. One may efficiently compute the brackets of these objects by starting with the fully-dressed version of the symplectic form~\eqref{Equation: Omega I fully-dressed}. From that equation it is simple to construct the reduced phase space corresponding to the Raychaudhuri constraint. Imposing $\tilde T=0$, the first term in the integral in~\eqref{Equation: Omega I fully-dressed} vanishes. Then $X$ just acts as a degenerate coordinate within the gauge orbit, and the constraints are completely first-class, so we can quotient by the gauge group by simply discarding the field $X$, ending up with
\be
\bm\Omega_{\text{red}} = \int_{I} \rd \mathrm{v} \sum_i \partial_{\mathrm{v}}\delta\tilde\varphi_i \wedge\delta\tilde\varphi_i + \sum_a\epsilon_a  \delta\omega_a\wedge\delta q_a,
\ee
which is a symplectic form for the dressed radiative fields $\tilde\varphi_i$, and the edge mode variables $\omega_a,q_a$.
One may read off the brackets
\begin{equation}
  \pb{\omega_a}{q_a} = \epsilon_a, \qquad
  \pb{\partial_\mv\tilde\varphi_i(\mv )}{\tilde\varphi_i(\mv ')} = \frac12\delta(\mv -{\mv '}).
\end{equation}
On the reduced phase space, $\tilde\tau$ is defined in terms of $\tilde\varphi_i$ via the constraint:
$\tilde\tau := -\sum_i(\partial_\mv \tilde\varphi_i)^2$; from this we can derive the brackets
\begin{align}
  \pb*{\tilde\tau(\mv )}{\tilde\varphi_i(\mv ')} &= - \delta(\mv -\mv ')\partial_\mv \tilde\varphi_i(\mv ),\\
  \pb*{\tilde\tau(\mv )}{\tilde\tau(\mv ')} &= -2 \partial_\mv\delta(\mv -\mv ')\tilde\tau(\mv ) -  \delta(\mv -\mv ')\partial_\mv \tilde\tau(\mv ).
\end{align}
As a consequence, $\tilde\tau$ acts as a stress tensor generating diffeomorphisms of the radiative fields $\tilde\varphi_i$. There is an overall minus sign appearing here, which accounts for the fact that these diffeomorphisms act on the left of the dressing field $V$, in contrast to gauge transformations which act on the right. We refer to them as `reorientations' and describe them further below.

The brackets of the dressed area may be obtained using~\eqref{area}. One has for example
\begin{equation}
  \pb{\tilde\Omega(\mv )}{q_0} = 8\pi\GN(\mv-1), \qquad
  \pb{\tilde\Omega(\mv )}{q_1} = 8\pi\GN\mv ,
\end{equation}
and
\begin{align}
  \pb{\tilde\Omega(\mv )}{\tilde\varphi_i(\mv ')} &= 8\pi\GN\int_0^1\dd{\mv ''}\Greens{\mv }{\mv ''} \pb{\tilde\tau(\mv '')}{\tilde\varphi_i(\mv ')}\\
  &= -8\pi\GN \Greens{\mv }{\mv '}\partial_{\mv'} \tilde\varphi_i(\mv' ).
\end{align}
Note in particular that $\tilde\Omega$ generates a \emph{non-local} transformation, because of the integration against the Green's function.

\subsubsection{Dirac brackets of bare fields}
\label{Section: dirac}

Alternatively one can use a gauge-fixing condition, and compute the resulting Dirac brackets. We will go to the gauge where $\beta=0$ and $v_a=a$, i.e.\ $X(\mv)=\mv$.

To this end, define
\begin{equation}
  \tilde T_{f} = \int_0^1 f \tilde T \dd\mv , \qquad X_h = \int_{0}^{1} h (X-\mv ) \dd\mv
\end{equation}
for functions $f, h:[0,1]\to\RR$. We impose $\tilde{T}_f=0$ and $X_h=0$. We have
\begin{equation}
  \pb{\tilde{T}_f}{X} = -f\partial_{\mv} X\onsheq -f,
\end{equation}
where $\onsheq$ denotes equality on shell of the constraints, so
\begin{align}
  \pb{\tilde T_f}{X_h} &\onsheq -\int_0^1h f\dd\mv .
\end{align}
Additionally,
\begin{equation}
  \pb{\tilde T_f}{\tilde T_g} \onsheq 0,\qquad \pb{X_h}{X_{h'}} = 0.
\end{equation}
We can straightforwardly invert this constraint algebra. We get a Dirac bracket in the following form:
\begin{align}
  \db{\mathcal{O}_1}{\mathcal{O}_2} &= \pb{\mathcal{O}_1}{\mathcal{O}_2} - \int_{0}^{1}\dd\mv  \qty\Big(\pb{\mathcal{O}_1}{\tilde T(\mv )}\pb{X(\mv )}{\mathcal{O}_2}- \pb{\mathcal{O}_2}{\tilde T(\mv )}\pb{X(\mv )}{\mathcal{O}_1})\\
  &\onsheq \pb{\mathcal{O}_1}{\mathcal{O}_2} + \int_{v_0}^{v_1}\dd{ v} \qty\Big(\pb{\mathcal{O}_1}{ T( v)}\pb{V(v)}{\mathcal{O}_2}- \pb{\mathcal{O}_2}{T(v)}\pb{V(v)}{\mathcal{O}_1})
\end{align}
where $\mathcal{O}_1,\mathcal{O}_2$ are any two observables, and these expressions should be evaluated at $\tilde{T}=X-\mv =0$, or equivalently $T=V-v=0$.

Since $q_a,\omega_a$ commute with both $T$ and $V$, their Dirac brackets with anything else reduce to Poisson brackets. Also, since the radiative fields $\varphi_i$ all commute with $V$, the Dirac brackets among them simply agree with the regular Poisson brackets. We only get a difference between Dirac and Poisson brackets when at least one of the observables depends on $\tau$. Using that $\tau$ satisfies $\pb{\tau(v)}{V(v')}=\partial_v V(v)\delta(v-v')$ and $\pb{\tau(v)}{T(v')}=-\delta(v-v')\partial_{v'}\tau(v')-2\partial_{v'}\delta(v-v')\tau(v')$, one finds
\begin{align}
  \db{\tau(v )}{\varphi_i(v ')} &= -\delta(v -v ')\partial_v \varphi_i(v ),\\
  \db{\tau(v )}{\tau(v ')} &= -2 \partial_{v}\delta(v -v ')\tau(v ) - \delta(v -v ')\partial_v \tau(v ).
\end{align}
Since we have fixed $V=v$, we have $\tau=\frac1{8\pi\GN}\partial_v^2\Omega$, from which the area brackets can be obtained.

\subsubsection{Isomorphism via the dressing map}

It should be clear that the brackets obtained by the above two approaches are isomorphic: the Dirac brackets of the bare observables are equal to the Poisson brackets of their dressed counterparts. Let us make this precise.

Let $\mathcal{O}$ be some (not necessarily gauge-invariant) kinematical observable, and suppose we fix the gauge such that dressing time agrees with background time, i.e. we set
\be
X(\mv) = \mv, \qquad V(v)= v,
\ee
as in Section~\ref{Section: dirac}. For the observable, this means restricting it to the subspace of phase space in which $X(\mv) = \mv$:
\begin{equation}
  \mathcal{O}_{\text{g.f.}} = \mathcal{O}|_{X(\mv)=\mv}.
\end{equation}
Conversely, any observable in this gauge-fixed subspace can be extended to a gauge-invariant observable on the full phase space by pushing forward its dependence on the fields through the dressing time. We call the map from gauge-fixed observables to gauge-invariant ones the `dressing map', and denote it $\Dmap$:
\begin{equation}
  \tilde{\mathcal{O}}  = \Dmap(\mathcal{O}_{\text{g.f.}}).
\end{equation}
For example:
\be
\mathcal{O}=\int_{v_0}^{v_1} \rd v f(v) \varphi_i(v) , \qquad
\mathcal{O}_{\text{g.f.}}=
\int_{0}^{1} \rd \mv f(\mv) \varphi_i(\mv),
\qquad
\Dmap(\mathcal{O}_{\text{g.f.}})
= \int_{0}^{1} \rd \mv f(\mv) \tilde{\varphi}_i(\mv).
\ee

The Poisson bracket of dressed observables agrees with the dressed Dirac bracket of the gauge-fixed observables:
\begin{equation}
  \pb{\Dmap(\mathcal{O}_1)}{\Dmap(\mathcal{O}_2)} = \Dmap(\db{\mathcal{O}_1}{\mathcal{O}_2}).
\end{equation}
This clearly holds for the fundamental degrees of freedom $\tau,\varphi_i,\omega_a,q_a$ by the previous explicit calculations, which implies that it holds for all observables formed from these degrees of freedom. Therefore, the dressing map $\Dmap$ provides an isomorphism between two Poisson algebras: the gauge-invariant observables equipped with the Poisson bracket, and the gauge-fixed observables equipped with the Dirac bracket.

\subsection{Diffeomorphism actions}
\label{Section: diffeomorphism actions}

There are three crucial ways in which diffeomorphisms act on the kinematical degrees of freedom of the null ray segment, which we call reparametrizations, reorientations, and dressed reparametrizations respectively -- see Table~\ref{Table: diffeomorphisms} for a summary of their properties. Each of these are important for different reasons: reparametrizations are the gauge transformations of the theory, reorientations are physical symmetries of the reference frame, and the charge generating dressed reparametrizations measures additional gauge-invariant degrees of freedom in the effective anomalous theory.

\begin{table}
  \centering
  \begin{tabular}{lcccc}\toprule
    & \small Action & \small Invariant fields & \small Generator & \small Algebra \\ \midrule
    {\small Reparametrization} &  $\delta_fV=f\partial_v V$ & $\tilde\tau,\tilde\varphi_i$ &$T_f = \int_{\Segment}\dd{v}fT$ & $\pb{T_f}{T_{f'}}=-T_{[f,f']}$\\\addlinespace
    {\small Reorientation} & $\delta_gV=g\circ V$ & $\tau,\varphi_i$ & $Q_g = \int_{I}\dd{\mv}g\tilde\tau$ & $\pb{Q_g}{Q_{g'}}=Q_{[g,g']}$\\\addlinespace
    {\small
      \begin{tabular}{@{}l@{}}Dressed \\ reparametrization
    \end{tabular}} &  $\delta_{\tilde f}V=\tilde f \circ V$ & $T,\tilde\varphi_i$ &$\tilde T_{\tilde f} = \int_{I}\dd{\mv}\tilde f \tilde T$ & $\pb*{\tilde T_{\tilde f}}{\tilde T_{\tilde f'}}=\tilde T_{[\tilde f,\tilde f']}$\\\addlinespace
  \end{tabular}
  \caption{Three kinds of diffeomorphism action are important for the gravitational degrees of freedom along the null ray segment: reparametrizations, reorientations, and dressed reparametrizations. These are each defined by whether they act on the left or right of the dressing time $V$, and which fields they leave invariant.}
  \label{Table: diffeomorphisms}
\end{table}

To derive the charges generating these diffeomorphisms, it will be useful to obtain some alternative expressions for the symplectic form~\eqref{Equation: Omega I}, that use either the original variables or the gauge invariant variables as follows.

Using~\eqref{Equation: left right maurer cartan} in~\eqref{Equation: Omega I} one finds
\begin{equation}
  \bm\Omega_{\Segment} = \delta\qty(-\int_{I}\dd{\mv}\tilde\tau\frac{\delta X}{\partial_\mv X} + \int_{\mathcal{I}}\dd{v}\sum_i \partial_v\varphi_i\delta\varphi_i)+ \sum_a \epsilon_a \delta\omega_a \wedge \delta q_a.
  \label{Equation: Omega I half-dressed}
\end{equation}
One may also change variables from $\varphi_i$ to $\tilde\varphi_i$ and use the identities 
\begin{equation}
  \delta \varphi \circ X =  \delta \tilde{\varphi} - \pa_\mathrm{v}\tilde{\varphi} \frac{\delta X}{\pa_{\mathrm{v}}X}  , \qquad
  (\pa_v \varphi)\circ X = \frac{\pa_{\mv} \tilde{\varphi}}{\pa_{\mv} X},
\end{equation}
to write this as  
\begin{equation}
  \bm\Omega_{\Segment} = \delta\qty(\int_{I}\dd{\mv}\qty\bigg(-\tilde T\frac{\delta X}{\partial_\mv X} + \sum_i \partial_\mv\tilde\varphi_i\delta\tilde\varphi_i))+ \sum_a \epsilon_a \delta\omega_a \wedge \delta q_a.
  \label{Equation: Omega I fully-dressed}
\end{equation}
Finally, one may change variables again from $\tilde T,X$ to $T,V$ to obtain
\begin{equation}
  \bm\Omega_{\Segment} = \delta\qty(\int_{\Segment} \dd{v}T\frac{\delta V}{\partial_v V} + \int_I\dd{\mv}\sum_i \partial_\mv\tilde\varphi_i\delta\tilde\varphi_i)+ \sum_a \epsilon_a \delta\omega_a \wedge \delta q_a.
  \label{Equation: Omega I half-dressed 2}
\end{equation}
One may expand out each of these expressions,  using the Maurer-Cartan zero curvature identities~\eqref{Equation: zero curvature} and~\eqref{Equation: zero curvature tilde}. For example, applying this to~\eqref{Equation: Omega I fully-dressed}, one finds
\begin{equation}
  {\bm\Omega}_{\Segment} = \int_0^1\dd\mv
  \qty(-\delta \tilde T \wedge\frac{\delta X}{\pa_\mv X}  - \frac12 \tilde{T}\left[\tfrac{\delta X}{\pa_\mv X},\tfrac{\delta X}{\pa_\mv X}\right]
  +\sum_i\partial_\mv \delta\tilde\varphi_i\wedge\delta\tilde\varphi_i)+ \tfrac{1}{2\pi}\sum_a \delta\omega_a \wedge \delta q_a.
  \label{Equation: Omega I unpacked}
\end{equation}

Let us now describe the various diffeomorphism actions and their charges.

\subsubsection{Reparametrizations}
\label{Subsection: reparametrizations}

The first action is a restriction of  the gauge transformations we have already been considering, which may be understood as \textbf{reparametrizations} of the fields as functions of the background time $v$. These gauge transformations leave invariant the dressed fields $\tilde\tau,\tilde\varphi_i$ and the edge mode variables $\omega_a,\eta_a$, but act non-trivally on $V$ as
\be
\delta_f V = f \pa_v V,
\label{Equation: V right action}
\ee
where $f$ is any smooth function. This corresponds to an infinitesimal version of the \emph{right action} on $V$, i.e.\ $F\triangleright V= V\circ F$. One may use the symplectic form derived above to confirm that the generator of this transformation is $T_f = \int_{\Segment}\dd{v}fT$, i.e.\ $I_{\delta_f}\bm\Omega_{\Segment}=-\delta T_f$ (assuming $\delta f=0$).
Note that this diffeomorphism action is an \emph{anti}morphism, meaning $[\delta_f,\delta_{f'}]= -\delta_{[f,f']}$. Hence, the charges satisfy the Poisson bracket
\be
\{ T_f, T_{f'} \} =
- T_{[f,{f'}]}.
\ee

As expected, the charge is the Raychaudhuri constraint integrated along $\Segment$. To show this, it is most straightforward to use~\eqref{Equation: Omega I half-dressed 2}, where the only relevant term is $\delta\qty(\int_{\Segment} \dd{v}T\frac{\delta V}{\partial_v V})$, which is a Kirillov-Kostant-Souriau (KKS) form on the cotangent bundle of the diffeomorphism group. The important point is that in this expression $T$ plays the role of the momentum conjugate to the left-invariant Maurer-Cartan form, which immediately implies that it generates the right action~\eqref{Equation: V right action} (according to the general theory of KKS forms). In more detail, one has that the symplectic potential $\int_{\Segment} \dd{v}T\frac{\delta V}{\partial_v V}$ is invariant under a finite reparametrization, since
\begin{equation}
    \int_{\Segment} \dd{v}T\frac{\delta V}{\partial_v V} \stackrel{F}\longrightarrow \int_{F^{-1}(\Segment)} \dd{v}\partial_v F\, T\circ F \qty(\frac{\delta V}{\partial_v V})\circ F = \int_{\Segment} \dd{v}T\frac{\delta V}{\partial_v V},
\end{equation}
where we used the fact that $\theta_{\text{diff}}=-\frac{\delta V}{\partial_v V}$ transforms like a vector, and that the integrand in the middle expression is the pullback through $F$. It follows that
\begin{equation}
    I_{\delta_f}\delta\qty(\int_{\Segment} \dd{v}T\frac{\delta V}{\partial_v V})\bm\Omega_{\Segment} = \underbrace{L_{\delta_f}\int_{\Segment} \dd{v}T\frac{\delta V}{\partial_v V}}_{=0} - \,\delta\qty(I_{\delta_f}\int_{\Segment} \dd{v}T\frac{\delta V}{\partial_v V}) =-\delta T_f,
\end{equation}
as required.

As an interesting aside, having carried out the split into the segment $\Segment$ and its complement $\SegmentBar$, one may now establish a slightly different notion of reparametrization, which has the same general form as above, but with different requirements on the properties of the function $f$. Indeed, suppose we only demand that $f$ is smooth \emph{inside} $\mathcal{I}$. Moreover, we assume that $f(v_a)=0$, and that this transformation acts trivially on $\bar V$, so that the transformation respects the split between $\mathcal{I}$ and $\bar{\mathcal{I}}$. Finally, we allow $f$ to be non-differentiable at $v_a$, so that this reparametrization acts non-trivially on the edge mode $q_a$:
\begin{equation}
  \delta_f q_a = \delta_f \eta_a - \delta_f \bar \eta_a = \lim_{\varepsilon\to 0}\epsilon_a\qty(\partial_v f(v_a-\varepsilon) - \partial_v f(v_a+\varepsilon)) = \partial_vf(v_a),
\end{equation}
where on the right-hand side $\partial_vf(v_a)$ is defined from inside $\Segment$. An example of such non-smooth reparametrizations are half-sided boosts of the segment, which are supposed to correspond to modular flow in a regularized theory, or indeed in full quantum gravity; it therefore seems important to ultimately account for them. One may use \eqref{Equation: Omega I} to confirm that
$I_{\delta_f} \Omega_{\Segment} = -\delta T_f$ (assuming $\delta f=0$), where the constraint is now
\begin{align}
  T_f
  = \int_{\Segment}\dd{v} f T - \sum_a \epsilon_a \omega_a \pa_vf_a  .
\end{align}
This is not the same constraint as we started out with. To take it seriously as a gauge constraint, one would have to treat the edge mode corner areas $\omega_a$ as sources of energy-momentum (as was done in~\cite{kirklin2024generalisedsecondlawsemiclassical}). To see this, one can write the constraint as the integral $T_f =\int_{\mathbb{R}} f(T \theta_{\mathcal{I}}- \omega_a \pa_v^2\theta_{\mathcal{I}}) \rd v $, where
$\theta_{\mathcal{I}}(v)$ is the characteristic function of the interval $\mathcal{I}$, such that $\theta_{\mathcal{I}}(v)=1$ if $v\in \mathcal{I}$ and $\theta_{\mathcal{I}}(v)=0$ if $v\in \bar{\mathcal{I}}$.
Here we see that the real value of having introduced the edge mode is that it allows one to consider the interval $\mathcal{I}$ independently of its complement and to allow for one-sided gauge transformations \cite{Donnelly:2016auv}. In quantum gravity, the momentum conjugated to the edge mode provides some kind of ``firewall'' \cite{Almheiri:2012rt}, i.e. a source for the constraints localized at the corner, which allows for the disentanglement of the interval and its complement. 

For completeness, let us note that one can split off the part of the half-sided transformation that acts on the edge modes. Indeed, the transformation
\be
\delta_{(\gamma_0,\gamma_1)}q_a =  \gamma_a, \qquad
\delta_{(\gamma_0,\gamma_1)} \varphi_i=\delta_{(\gamma_0,\gamma_1)}V= \delta_{(\gamma_0,\gamma_1)}\tau =\delta_{(\gamma_0,\gamma_1)} v_a=0
\ee
which leaves all the fields invariant except $\eta_a$, corresponds to a half-sided boost which changes the first derivative of $\bar{V}$ at the cut, but not $V$. Its charges are
\be
q_{(\gamma_0,\gamma_1)}=\gamma_1 \omega_1 -\gamma_0 \omega_0.
\ee

\subsubsection{Reorientations}

The next important diffeomorphism action is one corresponding to a \textbf{reorientation}~\cite{Hamette2021,Carrozza2022,Carrozza2024,Goeller:2022rsx} of the dressing time reference frame. It acts on the \emph{left} of the
dressing field $V:\mathcal{I}\to I$ without changing the bare radiative fields $\varphi$ or spin 0 stress tensor $\tau$.

In more detail, reorientations are labelled by a vector field $g(\mv)\pa_{\mv}$ on the reference interval $I$, and leave the bulk fields invariant 
\be \label{reor1}
\delta_g V = g\circ V \qq{or} \delta_g X = -g\partial_\mv X, \quad\mathrm{and}\quad 
\delta_g \tau = \delta_g \varphi_i = 0.
\ee
We shall assume reorientations do not act on the external embedding field $\bar X$, which implies that
\begin{equation}
  \delta_g v_a = \delta_g \bar X (a) = 0, \qquad \delta_g q_a = \delta_g \ln[\partial_vV(v_a)] = \partial_v g(a).
\end{equation}
So reorientations act non-trivially on the relative boost edge mode $q_a$. Also, the function $g$ must obey $g(0)=g(1)=0$, since we have defined $V$ such that $V(v_a)=a$ with $a=0,1$, so we must have $0 = \delta_g(a) = \delta_g V(v_a)=g(a)$. This means that $G=\exp g \pa_{\mv}$ is a diffeomorphism that preserves the reference interval $I=G(I)$. The reorientation symmetry group is therefore $\Diff(I)$. Since the reorientations preserve the  endpoints of the segment, it is natural to take $\delta_g \omega_a=0$.

To find the charge generating a reorientation, we can use $\bm\Omega_{\Segment}$ in the form~\eqref{Equation: Omega I half-dressed}. Since $\varphi_i$ is unaffected by the reorientation, we can focus on the terms involving $\tilde\tau, X$ and the edge modes.  One finds $I_{\delta_g}\bm\Omega_{\Segment} = -\delta Q_g$, where
\begin{align}
  Q_g &= \int_{0}^{1} g(\mv) \tilde{\tau}(\mv)  \rd \mv + \sum_a\epsilon_a\pa_{\mv}g(a)\omega_a = \int_0^1 \pa_{\mv}^2 g  \tilde{\Omega} \rd \mv
\end{align}
is the charge. The edge mode contribution is simple to confirm. The bulk integral over $\tilde\tau$ comes from the fact that $-\delta\qty(\int_{I} \dd{\mv}\tilde\tau\frac{\delta X}{\partial_\mv X})$ is again a KKS form, with $\tilde\tau$ now conjugate to the \emph{right}-invariant Maurer Cartan form $\delta X/\partial_\mv X$.

Thus, the reorientation charges are given by moments of the gauge-invariant area element.
This charge is gauge-invariant and doesn't vanish on-shell, so reorientations are physical symmetries (unlike reparametrizations).
Note also that the reorientation action is a morphism, meaning $[\delta_g,\delta_{g'}]
= \delta_{[g,g']}$. We therefore get the Poisson algebra
\be
\{Q_g, Q_{g'}\}= \delta_g Q_{g'} = Q_{[g,g']}.
\ee
Indeed, one may confirm this by noting that $\tilde\tau$ transforms as a rank 2 tensor with respect to the reorientation action:
\begin{align}
  \delta_g \tilde{\tau}
  = -(g\partial_\mv\tilde\tau+2\partial_\mv g \tilde\tau).
\end{align}
The action of reorientations on the area element may then be found with~\eqref{area} and $\delta_g\omega_a=0$ to be
\begin{align}
  \delta_g\tilde\Omega &=- \int_0^1\dd{\mv'}\Greens{\mv}{\mv'}(g\partial_{\mv'}\tilde\tau+2\partial_{\mv'} g \tilde\tau) = \int_0^1\dd{\mv'} \tilde\Omega\partial_{\mv'}^2[g(\mv'),\mathcal{G}(\mv,\mv')],
\end{align}
where
\begin{equation}
  [g(\mv'),\mathcal{G}(\mv,\mv')] = g(\mv')\partial_{\mv'}\mathcal{G}(\mv,\mv') - \mathcal{G}(\mv,\mv')\partial_{\mv'}g(\mv').
\end{equation}

\subsubsection{Dressed reparametrizations}

Finally, besides reparametrizations labelled by $f \in \mathrm{Vect}(\mathcal{I})$, and reorientations labelled by $g\in \mathrm{Vect}(I)$, we also will define \textbf{dressed reparametrizations}. These act on the left of $V$ (like reorientations), but leave the \emph{dressed} radiative fields $\tilde\varphi_i$ invariant (unlike reorientations, which leave the \emph{bare} radiative fields $\varphi_i$ invariant):
\be
\delta_{\tilde{f}} V = \tilde{f}\circ V, \qquad
\delta_{\tilde{f}} \tilde{\varphi}_i = 0
\ee
This means that the bare fields do transform:  $\delta_{\tilde{f}} \varphi_i = - \tilde{f}\circ V (\pa_v \varphi_i/ \pa_vV)$. This transformation corresponds to a field-dependent diffeomorphism $f= (\tilde{f}\circ V)/ \pa_v V $.
We shall also require dressed reparametrizations to preserve the bare stress tensor:
\be
\delta_{\tilde{f}} T = 0;
\ee
this guarantees that dressed reparametrizations have a gauge-invariant charge. One then finds that dressed reparametrizations act non-trivially on the dressed stress tensor $\tilde{T}$ as
\be
\delta_{\tilde f}\tilde T =- \left( \partial_\mv\tilde f \tilde T + 2\tilde f\partial_{\mv}\tilde T\right).
\ee
We shall also take the edge mode variables to be invariant under this transformation:
\begin{align} \label{drrepar2}
  \delta_{\tilde f} q_a = \epsilon_a(\delta_{\tilde f}\eta_a-\delta_{\tilde f}\bar\eta_a)= 0
  \qquad
  \delta_{\tilde f} \omega_a = \delta_{\tilde f} v_a
  = \delta_{\tilde f}\tilde{p}_a = 0.
\end{align}
As a consequence, dressed reparametrizations preserve the interval $I$, hence $\tilde f(0)=\tilde f(1)=0$.

To evaluate the charge, we can use the form~\eqref{Equation: Omega I fully-dressed} for $\bm\Omega_{\Segment}$, in which the only relevant term is $\delta\qty(-\int_{\Segment}\dd{\mv}\tilde T\frac{\delta X}{\partial_\mv X})$. Here, $\tilde T$ is the conjugate variable to the right-invariant Maurer-Cartan form. From this we can immediately deduce that the charge is
\begin{align}
  \tilde{T}_{\tilde{f}} &=\int_{0}^{1} \tilde{f}(\mv)\tilde T(\mv)\dd{\mv} = \int_{0}^{1} \tilde{f}(\mv) \qty\Big(\tilde\tau + \sum_i (\pa_{\mv}\tilde\varphi_i)^2 )   \rd \mv.
\end{align}
Like reorientations, dressed reparametrizations give a morphism of the algebra of diffeomorphisms $[\delta_{\tilde{f}}, \delta_{\tilde{f}}]=\delta_{[\tilde{f},\tilde{g}]}$. Therefore, the charges satisfy the algebra
\be
\{\tilde{T}_{\tilde{f}}, \tilde{T}_{\tilde{g}}\}
= \delta_{\tilde{f}}\tilde{T}_{\tilde{g}} =
\tilde{T}_{[\tilde{f},\tilde{g}]}.
\ee
These charges are gauge-invariant -- but they vanish on-shell, and hence do not generate physical symmetries.

\section{Effective null ray segments}
\label{Section: effective}

We have so far described the purely classical theory. In the quantum theory, fluctuations of the fields lead to extra contributions to the energy-momentum not yet accounted for. As discussed in~\cite{Ciambelli_2024}, these contributions produce anomalies in the quantum representations of diffeomorphisms. In this section, we will account for these anomalies at the \emph{classical} level, by formulating an effective theory of gravitational null ray segments.

We will start by describing in general terms how canonical quantization typically produces diffeomorphism anomalies, before explaining how such anomalies can be captured in the classical theory by appropriate deformations of the stress tensor and symplectic form. After giving some comments on the physical interpretation of these deformations, we will proceed with analyzing how they affect the structure of the theory. In particular, we will explain how they lead to modifications in the edge mode structure upon reduction to a segment, how they produce central charges in the algebras of the three diffeomorphism actions described in Subsection~\ref{Section: diffeomorphism actions}, and how gauge-invariance should be imposed in the presence of these anomalies.

\subsection{Canonical quantization and anomalies}
\label{Subsection: canonical anomalies}

Canonical quantization is the replacement of Poisson brackets of classical observables by the commutators of corresponding quantum operators:
\begin{equation}
  \pb*{\mathcal{O}_1}{\mathcal{O}_2} = \mathcal{O}_3
  \quad \longrightarrow \quad
  \comm*{\hat{\mathcal{O}}_1}{\hat{\mathcal{O}}_2} = -i\hbar\hat{\mathcal{O}}_3.
\end{equation}
But this replacement is typically only exact for a restricted set of observables (usually the ones linear in the fields) -- for all the others there are corrections on the right-hand side. In QFT, one must normal-order operators to make sense of them, which is a significant source of such corrections.

The classical algebra of diffeomorphism generators is isomorphic to the Lie algebra of the vector fields they represent. For example, we have seen that
\begin{equation}
  \pb{T_f}{T_g} = -T_{[f,g]},
  \label{Equation: tree-level algebra}
\end{equation}
where $[f,g] = f\partial_v g - g\partial_v f$ is Lie bracket.
In the quantum theory, these generators are normal-ordered $T_f\to \normord{T_f}$, which modifies the stress tensor algebra as follows:
\begin{equation}
  \frac1{i\hbar}\comm{\normord{T_f}}{\normord{T_g}} = \normord{T_{[f,g]}} - \frac{c\hbar}{48\pi}\int \dd{v}(f\partial_v^3g-g\partial_v^3 f).
  \label{Equation: T algebra anomaly}
\end{equation}
The additional term on the right is the unique non-trivial central extension of~\eqref{Equation: tree-level algebra}, and the resulting structure is known as the Virasoro algebra.\footnote{ To get the standard presentation of the Virasoro algebra in terms of modes $L_n$, one can define a periodic coordinate $\theta$ via a mapping $v:[0,2\pi]\to\mathbb{R}$ such that 
\begin{equation}
  v(\theta)=\tan\frac{\theta}{2}, \qquad e^{i\theta(v)}=\frac{1+iv}{1-iv},
\end{equation}
which satisfies $\partial_\theta v=\frac12(1+v^2)$ and has Schwarzian derivative $\Schwarzian{v}{\theta}=\frac12$. Then we can consider the charges corresponding to the diffeomorphisms generated by 
\begin{equation}
  e^{in\theta(v)}\partial_\theta = e_n(v)\partial_v, \qq{where} e_n(v)= \tfrac12(1+iv)^{1+n}(1-iv)^{1-n}.
\end{equation}
Setting $L_n=\normord{T_{e_n}}$, and using natural units $\hbar=1$, one finds
\begin{equation}
  [L_m,L_n] = (m-n)L_{n+m} + \frac{c}{12}(m^3-m)\delta_{n+m,0},
\end{equation}
which agrees with e.g.~\cite{Polchinski_1998}.} The uniqueness of this central extension guarantees that any consistent quantization procedure will produce a term of this kind. The parameter $c$, whose value is typically non-zero and depends on the details of the quantization, is called the `central charge'.

One may exponentiate the quantum diffeomorphism generator $\normord{T_f}$ to obtain the unitary operator $U[F] = \exp(i\normord{T_f}/\hbar)$ that carries out the finite diffeomorphism $F=\exp(f\partial_v)$. One then finds that the quantum transformation law from the stress tensor differs by a term proportional to the central charge:
\begin{equation}
  U[F]\normord{T(v)}U[F]^{\dagger} = (\partial_vF)^2\normord{T(F(v))} - \frac{c\hbar}{24\pi}\Schwarzian{F}{v}.
  \label{Equation: T(v) anomaly}
\end{equation}
Here
\begin{equation}
  \Schwarzian{F}{v} = \partial_v\qty(\frac{\partial_v^2F}{\partial_vF}) - \frac12\qty(\frac{\partial_v^2F}{\partial_vF})^2
\end{equation}
is the so-called Schwarzian derivative of $F$.

There is therefore a discrepancy between the algebras of diffeomorphisms in the classical and quantum theories. This mismatch does not necessarily go away in the classical limit, because the Schwarzian derivative can be arbitrarily large, so there are always diffeomorphisms $F$ for which the second term on the right-hand side of~\eqref{Equation: T(v) anomaly} is non-negligible. For this reason, the `canonical quantization' described above is not really the `correct' quantization of the classical degrees of freedom, because its classical limit does not recover the original classical theory. For the classical gauge constraint $T=0$ this is particularly evident, since the presence of a quantum anomaly~\eqref{Equation: T algebra anomaly} means it is impossible to impose $\normord{T}=0$ on quantum states. On the other hand, since the anomaly is central, one can consistently define gauge invariant operators to be operators that commute  with $\normord{T}$. We will describe the classical version of this below.

Two obvious remedies are possible. First, one can seek a different quantization procedure for which $c=0$ at the quantum level, so that the quantum theory really does reproduce the original classical theory in the classical limit. This is a subtle game, because there are many properties one typically would want the quantum theory to obey, such as unitarity, boundedness, etc., and these properties are not necessarily possible to guarantee at $c=0$. The alternative is to take the theory obtained by canonical quantization at its word, and then to \emph{replace} the original classical theory with the classical limit of this quantum theory.

In what follows we will pursue the latter procedure. We will refer to the original classical theory of Section~\ref{Section: classical gravity} as the \emph{tree-level} theory, and its replacement as the \emph{effective} theory.
\begin{figure}
  \centering
  \begin{tikzpicture}[]
    \begin{scope}[shift={(0,0.2)}]
      \node at (0,0) {tree-level};
      \node at (0,-0.5) {\small$c=0$};
      \node at (5,0) {quantum};
      \node at (5,-0.5) {\small$c\ne0$};
      \node at (10,0) {effective};
      \node at (10,-0.5) {\small$c\ne0$};
    \end{scope}

    \draw[-Latex] (1.2,0) -- (3.8,0) node[midway, above, align=center, text width=3cm,font=\footnotesize\linespread{0.9}\selectfont] {`canonical quantization'};
    \draw[Latex-,shift={(0,0.3)}] (6.2,0) .. controls (7,0.5) and (8,0.5) .. (8.8,0) node[midway, above, align=center, text width=3cm,font=\footnotesize\linespread{0.9}\selectfont] {actual quantization};
    \draw[-Latex,shift={(0,-0.3)}] (6.2,0) .. controls (7,-0.5) and (8,-0.5) .. (8.8,0) node[midway, above=1.5pt, align=center, text width=1.5cm,font=\footnotesize\linespread{0.9}\selectfont] {classical limit};
  \end{tikzpicture}
  \caption{A na\"ive canonical quantization of the tree-level theory results in a non-zero central charge $c\ne0$. The classical limit of this quantized theory is the \emph{effective theory}. It replaces the original tree-level description.}
  \label{Figure: tree level vs effective}
\end{figure}
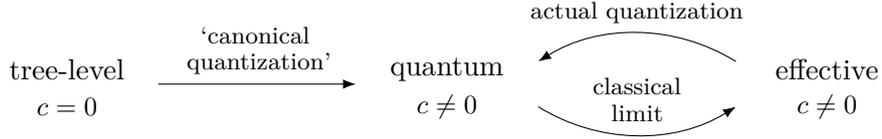
Going to the effective theory is not an ad hoc prescription. The effective theory accounts for additional contributions to the classical equations of motion from a 1-loop effective action (i.e.\ the result of integrating out quantum fluctuations around a classical saddlepoint in a path integral). Such contributions capture important physics, for example Hawking radiation; indeed, at a horizon, the anomaly encodes (as we'll discuss later below) the energy flux of thermal radiation at the Hawking temperature~\cite{Candelas:1977zz,Robinson_2005,PhysRevD.13.2720}. The effective theory thus is capable of describing a (near-)classical limit of an evaporating black hole, whereas the original tree-level description is not.

\subsection{Anomaly counterterms}
\label{Subsection: anomaly counterterms}

In order to properly account for an anomaly at the level of the classical theory, we need to deform both the stress tensor and the symplectic form:
\begin{equation}
  T\longrightarrow T^c = T + c\hbar T_{\text{an}}, \qquad \bm\Omega \longrightarrow \bm\Omega^c = \bm\Omega + c\hbar \bm\Omega_{\text{an}},
\end{equation}
where $(T, \bm{\Omega})$
are the stress tensor and symplectic form respectively of the tree-level theory given by 
\begin{align}
  T &= \tfrac1{8\pi \GN}\qty\big(\partial_v^2\Omega-\beta\partial_v\Omega) + \sum_i(\partial_v\varphi_i)^2
  = \tau + \sum_i(\partial_v\varphi_i)^2, \\
  \bm\Omega &= \int_{\Ray}\dd{v}\qty\Big(\tfrac1{8\pi\GN} \delta \Omega \wedge\delta\beta + \sum_i \partial_v(\delta\varphi_i)\wedge\delta\varphi_i).
\end{align}
The first deformation $T\to T^c$ is required in order to have a constraint with an anomalous transformation law:
\begin{equation}
  F\triangleright T^c = (\partial_v F)^2 T^c\circ F - \frac{c\hbar}{24\pi} \Schwarzian{F}{v}.
  \label{Equation: anomalous effective T}
\end{equation}
The original Raychaudhuri constraint satisfies this equation when $c=0$, but we want to allow for $c\ne 0$. Having changed $T\to T^c$, the second deformation $\bm\Omega\to\bm\Omega^c$ is required in order to ensure that $T^c$ is still the generator of gauge transformations, i.e.\ $I_{\delta_f} {\bm\Omega}_{c} =- \delta T^c_f$ where $T^c_f  = \int_{\Ray}\dd{v} fT^c$.
This suffices to guarantee that the corresponding charge algebra is centrally extended in the appropriate way:
\be
\{T^c_f, T^c_g\} = \delta_f T^c_g = - T^c_{[f,g]} + \frac{c\hbar}{48 \pi}  \int_{\Ray}\dd{v} (f\pa_v^3g-g\pa_v^3f).
\ee

We choose to explicitly include factors of $\hbar$ in this paper. One could use natural units in which $\hbar=1$, but we would like to emphasize the quantum origin of the anomaly. Note that including the $\hbar$ factors means $c$ is a dimensionless quantity. 

Let us now describe the precise forms of the two anomaly counterterms $T_{\text{an}}$, $\bm\Omega_{\text{an}}$, and show that they have the desired properties.

\subsubsection{Stress tensor}

To satisfy~\eqref{Equation: anomalous effective T}, we must have $F\triangleright T_{\text{an}} = (\partial_v F)^2  T_{\text{an}}\circ F - \frac{1}{24\pi} \Schwarzian{F}{v}$. Quite remarkably, as was already noticed in \cite{Ciambelli_2024}, we can construct this anomalous stress tensor purely form the spin $0$ momenta $\beta$ as 
\begin{equation}
  T_{\text{an}} := -\frac{1}{24\pi}\qty(\partial_v\beta - \frac12\beta^2) = -\frac{1}{24\pi}\Schwarzian{V}{v}.
  \label{Tanom}
\end{equation}
Thus, the counterterm is proportional to the Schwarzian derivative of the dressing time. Either from the anomalous transformation \eqref{varOb} of $\beta$ or from the chain rule for the Schwarzian, we get that $T_{\text{an}}$ transforms in the appropriate way.

This is the simplest possible choice of stress tensor counterterm, and the most convenient for our purposes, but it is worth pointing out that there are other possible choices. For example, we could substitute $\beta \to \beta +\alpha \theta$ into \eqref{Tanom}, where $\alpha$ is an arbitrary constant, and $\theta=\frac12\partial_v\ln\Omega$ is the expansion; this then gives a $T_{\text{an}}$ constructed purely from the spin 0 sector and consistent with the required anomalous transformation law. This choice of counterterm would be proportional to the Schwarzian of other possible times, as discussed in \cite{Ciambelli_2024}. In this paper, we will just use \eqref{Tanom}. We will discuss this point further in the Conclusion.

Therefore, in the effective theory, the total Raychaudhuri constraint is corrected as follows:
\begin{align}
  T\to T^c&=\tfrac1{8\pi\GN}\qty\big(\partial_v^2\Omega-\beta\partial_v\Omega) - \tfrac{c\hbar}{24\pi}\qty\big(\partial_v\beta-\tfrac12\beta^2) + \sum_i(\partial_v\varphi_i)^2\\
  &= \tau  + \sum_i(\partial_v\varphi_i)^2 - \frac{c\hbar}{24\pi} \Schwarzian{V}{v}.
  \label{Equation: overall Raychaudhuri}
\end{align}

\subsubsection{Symplectic form}
We want~\eqref{Equation: overall Raychaudhuri} to be the generator of a diffeomorphism, which implies that an anomalous counterterm in the symplectic form is also required. To construct this deformation, let us introduce a `boost field' $\eta(v)$ which satisfies $
\partial_v\eta = \beta,
$
and which
transforms anomalously under primed diffeomorphisms as
\begin{equation}\label{etaaction}
  F\triangleright \eta = \eta\circ F + \ln\partial_v F,
\end{equation}
Inside the segment $\Segment$, we define $\eta$ to be related to the dressing time as follows
\begin{equation}
  \eta|_{\Segment} = \ln\partial_v V.
  \label{Equation: k inside}
\end{equation}

Given the boost field we can consider the following anomalous contribution to the symplectic structure
\be
{\bm\Omega}_{\text{an}} :=
\frac{1}{48\pi} \int_{\Ray} \partial_v(\delta \eta) \wedge \delta  \eta \dd{v}.
\ee
For an infinitesimal gauge transformation
\begin{equation}
  \delta_f\eta = f\partial_v \eta + \partial_v f,
  \label{Equation: eta transformation}
\end{equation}
one may confirm that (assuming $\delta f=0$ and the gauge falloff conditions~\eqref{Equation: diffeo falloffs})
\begin{equation}
  -I_{\delta_f} {\bm\Omega}_{\text{an}} = -\delta\qty(\frac{1}{24\pi}\int_{\Ray}f \qty(\partial_v^2\eta-\frac12(\partial_v\eta)^2)\dd{v}) = \delta\qty(\int_{\Ray}f T_{\text{an}}\dd{v}).
\end{equation}
As a consequence, the anomalously deformed symplectic structure
\begin{align}
  \bm{\Omega}^c &= \bm{\Omega} + c\hbar{\bm\Omega}_{\text{an}}
  = \int_{\Ray}\dd{v}\qty\Big(\tfrac1{8\pi\GN} \delta \Omega \wedge\delta\beta + \tfrac{c\hbar}{48\pi}\partial_v(\delta\eta)\wedge\delta\eta + \sum_i \partial_v(\delta\varphi_i)\wedge\delta\varphi_i)
\end{align}
satisfies
\begin{equation}
  -I_{\delta_f}\bm\Omega^c = \delta T_f^c.
\end{equation}
In other words, this symplectic form is such that $T_f^c$ is the charge generating the diffeomorphism $\delta_f$, as required.

\subsection{Anomalous area and the effective Raychaudhuri equation}
\label{Subsection: anomalous area}

Let us now give some comments on the physical interpretation of the deformation of the theory outlined above.

The deformed symplectic form can be written in terms of an `anomalous area element'
\begin{equation}
  \Omega_c := \Omega - \frac{c\hbar\GN}{6}\eta
\end{equation}
as
\begin{align}
  \bm{\Omega}^c & = \int_{\Ray}\dd{v}\qty\Big(\tfrac1{8\pi\GN} \delta \Omega_c \wedge\delta\beta + \sum_i \partial_v(\delta\varphi_i)\wedge\delta\varphi_i).
\end{align}
In other words, $\bm\Omega^c$ can be obtained by the simple substitution $\Omega\to \Omega_c$ in the tree-level symplectic form $\bm\Omega$. Unlike the ordinary area element $\Omega$, this anomalous area element $\Omega_c$ transforms anomalously:
\begin{equation}
  \delta_f\Omega_c = f\partial_v \Omega_c - \frac{c \hbar\GN }{6} \partial_v f.
  \label{Equation: Omega c transformation}
\end{equation}
This action can be exponentiated into
\be
F\triangleright \Omega_c = \Omega_c \circ F -\frac{c\hbar\GN}{6} \ln\pa_v F.
\ee
Thus, the deformation from the tree-level theory to the effective one can alternatively be understood as the introduction of an anomalous transformation law for the area.

On the other hand, a different anomalous area element seems to be more physically relevant when interpreting the effective Raychaudhuri constraint $T^c$. Indeed, defining
\begin{equation}
  \Omega_{2c} := \Omega- \frac{c\hbar\GN}{3} \eta,
\end{equation}
one has
\be
T^c=\frac{1}{8\pi G}
(\pa_v-\beta)\pa_v \Omega_{2c} + T^{\text{rad}}_c, \qq{where} T^{\text{rad}}_c = - \frac{c\hbar}{24\pi} (\pa_v \eta)^2 + \sum_i (\pa_v \varphi_i)^2.
\ee
Comparing this to the tree-level constraint
\be
T=\frac{1}{8\pi G}
(\pa_v-\beta)\pa_v \Omega + T^{\text{rad}}, \qq{where} T^{\text{rad}} = \sum_i (\pa_v \varphi_i)^2,
\ee
one observes that $\Omega_{2c}$ evolves in the effective theory similarly to how $\Omega$ evolves in the tree-level theory, except that there is an additional kinetic term for the boost field $\eta$ in the deformed radiative stress energy $T^{\text{rad}}_c$.

The sign of $c$ appears to be rather important. If $c$ is negative, then this additional kinetic term has the same sign as that of the radiative fields $\varphi_i$, and so $\Omega_{2c}$ has the same kinds of dynamical properties as $\Omega$ in the tree-level theory -- for example it is subject to focusing, and one recovers a second law of area increase for $\Omega_{2c}$. On the other hand, if $c$ is positive, the new kinetic term for $\eta$ has the \emph{opposite sign} to that of the radiative fields, contributing to a negative energy flux across the surface. 

In a unitary quantum theory one usually expects $c$ to be positive. This reproduces the standard situation for quantum fields near a horizon: the effective negative energy flux generated by Hawking radiation allows for the shrinking of $\Omega_{2c}$, corresponding physically to horizon evaporation.

Even if $c$ is negative, such that $\Omega_{2c}$ is consistent with focusing and the area law $\Delta \Omega_{2c} >0$, the ordinary area element then only needs to obey
\be
\Delta\Omega > \frac{c\hbar\GN}{3}\Delta \eta.
\label{Equation: Omega GSL anomalous}
\ee
In principle, if $\Delta\eta$ is sufficiently large, one observes that the non-anomalous area $\Omega$ can macroscopically decrease, violating the tree-level area theorem.\footnote{It is interesting to note the resemblance between~\eqref{Equation: Omega GSL anomalous} and the modification to the generalized second law in~\cite{kirklin2024generalisedsecondlawsemiclassical}.}

Let us finally comment on another interpretation of the deformed constraint. As will be explained shortly, given a configuration of the tree-level stress tensor $T$, one can always construct a scalar $V_T$, called the Hill potential, such that $T(v) =\frac{c\hbar}{24\pi} \Schwarzian{V_T}{v}$. The construction of this scalar, detailed below follows from solving the Hill's equation for the Hill prepotentials associated with $T$. This $V_T$ is defined up to a  PSL$(2,\mathbb{R})$ M\"obius transformation:
\be
V_T \to \frac{\alpha V_T + \beta}{\gamma V_T+\delta}, \qq{with} \alpha\delta-\beta\gamma=1.
\label{Equation: mobius}
\ee
Therefore the deformed constraint $T^c=0$, hence $T= \frac{c\hbar}{24\pi}\{V,v\}$ can be written as 
\be 
\{V,v\}= \{V_T,v\}
\ee 
which means that the Hill's potential $V_T$ must be related to the dressing time $V$ by a PSL$(2,\mathbb{R})$ M\"obius transformation (since such transformations are the only ones which preserve the Schwarzian derivative). Thus, $V$ can be understood as a Hill potential for $T$.

By including the anomalous counterterm in the effective constraint, we are in this way effectively allowing for configurations which are off-shell of the tree-level constraint, absorbing them into the Hill potential $V=V_T$. The source term $\frac{c\hbar}{24\pi} \Schwarzian{V}{v}$ corresponds to the parametrization of a coadjoint orbit of the Virasoro group. This is related to the `extra particle' interpretation of QRFs~\cite{CastroRuiz2025}, where an additional degree of freedom (the eponymous `extra particle'), which carries an irreducible representation of the symmetry group, is used to keep track of the charges of a system. The extra particle deforms the gauge constraint by contributions from this additional degree of freedom, and we do the same here with the deformation $T_{\text{an}}$. The QRF extra particle lives in an irreducible representation of the gauge group; here we  similarly have that $V$ may be understood as having a phase space constructed from a coadjoint orbit of the Virasoro group~\cite{Alekseev:1988ce,Witten:1987ty,Cotler:2018zff,nair2024notecoherentstatesvirasoro,Stanford:2017thb}. After quantization, this becomes an irreducible representation of Virasoro, parametrized by the central charge $c$. Note that here this is not the most general possible irreducible representation -- there are other ones parametrized by numbers such as the vacuum value $b_0$ of the zero mode of the stress tensor, as well as the winding number $n$. Here we have chosen the \emph{minimal} irreducible representation, in the sense that it has the largest isotropy group -- in this case, $\SLTR$.

For completeness, let us give some details on how one may construct the Hill potential $V_T$, given $T$. We follow the presentation of \cite{Blau:2024owj}: Given a stress tensor $T$ and a central charge $c$ for this stress tensor, the Hill prepotentials are defined as solutions $\Phi$ of Hill's equation, which is the second order equation $\partial_v^2{\Phi}+ \frac{12\pi}{c\hbar} T \Phi=0$.

Given $T$ we can find two linearly independent solutions of Hill's equation $(\Phi,\Phi')$ which are $-1/2$ densities. To ensure linear independence we ask for the Wronskian condition $\Phi'\pa_v  \Phi- \Phi \pa_v  \Phi'=1$.
Given two such solutions, we can define a scalar function by taking the ratio
$V_T= \frac{\Phi'}{\Phi}$ which is such that \be \pa_v V_T = \frac{1}{\Phi^2}
\quad\mathrm{and}\quad
\frac{c\hbar}{24\pi}\Schwarzian{V_T}{v}= T.
\ee
 The monodromy of $V$ is a PSL$(2,\mathbb{R})$ transformation such that
$V_{T+}=\frac{\alpha V_{T-}+\beta}{\gamma V_{T-}+\delta}$ where $V_{T\pm} = V_T(\pm \infty)$. This follows from having that $T(+\infty)=T(-\infty)$.
The two Hill potentials can be obtained from $V_T$ as
\be
\Phi = \frac{1}{\sqrt{\pa_v V_T}}, \qquad
\Phi' = \frac{V_T}{\sqrt{\pa_v V_T}}.
\ee
It is interesting to note that the vector fields which preserve $T$ are given by
\be
f_{(p,q,r)}
= p \frac1{\pa_v V_T }  +
q \frac{V_T}{\pa_v V_T } + r \frac{V_T^2}{\pa_v V_T }.
\ee
These precisely generate the PSL$(2,\RR)$ transformations~\eqref{Equation: mobius}.

\subsection{Reduction to the segment}\label{externalD}

We will now show how to restrict the deformed phase space to the segment $\Segment\subset\Ray$.
To do so, we will follow the strategy employed in Section \ref{edgem}: we assume that the variations of $(\tau, \beta ,\varphi_i)$ outside $\mathcal{I}$ are gauge transformations (recall that $\tau$ is the spin zero stress tensor, $ 8 \pi \GN \tau= (\pa_v-\beta) \pa_v \Omega$).

The end result will be to write the total symplectic form as
\begin{equation}
  \mathbf{\Omega} = \mathbf{\Omega}_{\Segment} + \mathbf{\Omega}_{\text{edge}} + \mathbf{\Omega}_{\SegmentBar},
  \label{Equation: segment reduced anomalous symplectic form decomposition}
\end{equation}
where
\begin{align}
  \mathbf{\Omega}_{\Segment} &= \delta\qty(\int_{\mathcal{I}}\dd{v}\qty(\tau_c \frac{\delta V}{\pa_v V} + \sum_i \partial_v\varphi_i\delta\varphi_i)) ,\label{Equation: anomalous Omega I}\\
  \mathbf{\Omega}_{\text{edge}}&=\sum_a\epsilon_a\delta(\omega_a-\tfrac{c\hbar}{48\pi}\eta_a)\wedge\delta q_a,\\
  \mathbf{\Omega}_{\SegmentBar} &= \frac12\int_{\SegmentBar}\dd{v}\delta T^c\wedge\frac{\delta \bar V}{\partial_v \bar V} +\frac{c\hbar}{48\pi}\sum_a\epsilon_a\beta_a\delta\bar\eta_a\wedge\delta v_a
\end{align}
are the symplectic forms for the segment, the edge modes, and the complement respectively.\footnote{Note that in Section~\ref{Section: classical gravity} we included the edge modes and bulk of $\Segment$ together in $\bm\Omega_{\Segment}$; here we find it more convenient to split them up as in~\eqref{Equation: segment reduced anomalous symplectic form decomposition}.} Here $T^c=T+c\hbar{T}_{\text{an}}$ is the anomalous Raychaudhuri constraint, and we have defined
\begin{align}
  \tau_c &= \tau - \frac{c\hbar}{48\pi}\partial_vV \pa_v\qty(\frac{\pa_v \eta}{\partial_vV})
  = \partial_vV \partial_v\qty(\frac{\partial_v(\tfrac1{8\pi\GN}\Omega - \frac{c\hbar}{48\pi}\eta)}{\partial_v V})
  = \frac{\partial_vV}{8\pi\GN}\partial_v\qty(\frac{\partial_v\Omega_c }{\partial_v V}),
\end{align}
which replaces $\tau$ in the effective theory as the conjugate variable to the diffeomorphism group Maurer-Cartan form $\delta V/\partial_vV$.

The decomposition~\eqref{Equation: segment reduced anomalous symplectic form decomposition} involves a specific distribution of
the boundary contributions at $v=v_a$ among the three terms.
The chosen decomposition is justified for the following reasons. First, the symplectic forms $\Omega_{\Segment},\Omega_{\text{edge}}$ for $\Segment$ and the edge modes can be obtained from their tree-level counterparts by the straightforward replacement of the area element by the anomalous area element, $\Omega\to\Omega_c$. Second, the boundary term in $\Omega_{\SegmentBar}$ depends only on boundary values of $\bar{V}$ and its derivatives.\footnote{$\Omega_{\SegmentBar}$ depends on $\beta_a$, but $\beta_a$ itself can be written as a function of $\bar{V}$ and the background configuration $\bg\beta$; indeed, we have $\beta_a = \left.\qty(\partial_v\bar V\bg\beta + \frac{\partial_v^2\bar V}{\partial_v\bar V})\right|_{v_a}$.} And last but not least, each of the terms in~\eqref{Equation: segment reduced anomalous symplectic form decomposition} is a \emph{closed} two-form and hence can serve as a symplectic form for its respective degrees of freedom. For $\Omega_{\Segment},\Omega_{\text{edge}}$ this is manifest; for $\Omega_{\SegmentBar}$ it is less obvious, but will be confirmed below.

In the tree-level theory of Section~\ref{Section: classical gravity}, in the complement of $\Segment$ we went on-shell of the constraint, $T|_{\SegmentBar}=0$. We shall avoid explicitly doing so in the effective theory, because in the presence of an anomaly the constraints are second class, meaning $T^c|_{\SegmentBar}=0$ is not a gauge-invariant condition; we shall describe the consequences of this more in Subsection~\ref{Subsection: anomalous constraint}. In any case, we shall not need to set $T^c|_{\SegmentBar}=0$ to proceed.

Let us now show how to derive~\eqref{Equation: segment reduced anomalous symplectic form decomposition}. To formalize that the field variations $\delta(\tau,\beta,\varphi_i)$ are indistinguishable from gauge transformations in $\SegmentBar$, we shall take the field configurations in $\SegmentBar$ to be finite gauge transformations of some \emph{fixed} background field configurations denoted $(\bg{\tau},\bg\beta,\bg\varphi_i)$, satisfying $\delta(\bg{\tau},\bg\beta,\bg\varphi_i)=0 $. The fields outside $\mathcal{I}$ are therefore given by:
\begin{equation}
  \begin{gathered}
    \tau|_{\SegmentBar} = (\partial_v \bar V)^2 \bg{\tau}\circ \bar V, \qquad \varphi_i|_{\SegmentBar} = \bg\varphi_i\circ \bar V,\\
    \beta|_{\SegmentBar} = (\partial_v \bar V)\bg \beta\circ\bar V + \frac{\partial_v^2\bar V}{\partial_v \bar V}
    \label{Equation: background + bar V}
  \end{gathered}
\end{equation}
Here, we take the gauge transformation to be given by the dressing field $\bar V$ which maps the external intervals $[v_a, \epsilon_a \infty[$ to the reference intervals $[a, \epsilon_a \infty[$.

Let us emphasize that we are only introducing these external background configurations in order to make the following calculations explicit. Crucially, the final phase space and symplectic form for the segment \emph{do not} depend on the particular choice of external background field configurations.

It is important to note that although the variations of the fields $\tau,\beta$ are taken to be indistinguishable from gauge transformations outside $\Segment$, the same is \emph{not} necessarily true of $\Omega,\eta$. Indeed, $\Omega$ is related to $\tau$ by a second order differential equation, and $\eta$ is related to $\beta$ by a first order differential equation, and this means there are three additional constants of integration (in each of the disconnected regions $[v_a,\epsilon_a\infty[$) that parametrize the configurations of $\Omega,\eta$ outside $\Segment$. It turns out that these constants of integration are precisely related by continuity to the edge modes considered previously.

In particular, one finds
\begin{align}
  \Omega(v)|_{[v_a,\epsilon_a\infty[} = 8\pi\GN\left(\omega_a + \tilde p_a(\bg V-a) + \bg{\Omega}\right) \circ \bar{V}(v)
  \label{Equation: omega outside}
\end{align}
where $\bg\Omega$ is a background area element configuration satisfying $\bg\Omega(a)=\partial_v\bg\Omega(a)=0$ and $\bg V$ is a background dressing time configuration that satisfies $\bg\beta=\frac{\partial_v^2\bg V}{\partial_v\bg V}$. Similarly, we have that 
\begin{equation}
  \eta|_{[v_a,\epsilon_a\infty[} = \bg{\eta}\circ\bar V + \ln\partial_v\bar V + q_a,
  \label{Equation: eta outside}
\end{equation}
where $\bg\eta$ is a background configuration satisfying $\bg\eta(a)=0$ and $\partial_v\bg\eta=\bg\beta$. Recall that the edge modes contributing here are defined as
\begin{equation}
  \omega_a = \frac{\Omega(v_a)}{8\pi\GN},\qquad \tilde p_a = \frac1{8\pi\GN} \frac{\partial_v\Omega(v_a)}{\partial_vV(v_a)}, \qquad q_a = \eta_a-\bar\eta_a.
\end{equation}
One can confirm~\eqref{Equation: omega outside} and~\eqref{Equation: eta outside} by evaluating these expressions and their derivatives at $v=v_a$.

In Section~\ref{Section: classical gravity}, we showed that the tree-level symplectic potential under these assumptions reduces to
\begin{equation}
  \Theta = \int_{v_0}^{v_1}\dd{v}\qty(\tau\frac{\delta V}{\partial_v V} + \sum_i\partial_v\varphi_i\delta\varphi_i) + \sum_a \epsilon_a \omega_a\delta q_a + \int_{\SegmentBar}\dd{v} \frac{\delta \bar{V}}{\pa_v \bar{V}} T,
  \label{Equation: Theta tree level}
\end{equation}
where $T$ is the tree-level Raychaudhuri constraint. The final term here follows from substituting $f=\frac{\delta \bar{V}}{\partial_v\bar{V}}$ into~\eqref{Equation: Theta a}.
  Taking an exterior derivative of~\eqref{Equation: Theta tree level} yields the tree-level symplectic form
\begin{equation}
  \bm\Omega = \delta\qty(\int_{v_0}^{v_1}\dd{v}\qty(\tau\frac{\delta V}{\partial_v V} + \sum_i\partial_v\varphi_i\delta\varphi_i)) + \sum_a \epsilon_a \delta\omega_a\wedge\delta q_a + \delta\qty(\int_{\SegmentBar}\dd{v} \frac{\delta\bar V}{\partial_v\bar V} T).
  \label{Equation: tree level symplectic form segment}
\end{equation}
It will be useful to rewrite the last term. First, note that $T= (\pa_v\bar{V})^2 \bg{T}\circ \bar{V}$ (with $\bg{T}$ equal to $T$ evaluated on the background configuration) implies
\begin{align}
  \delta T &
  =\left(\frac{\delta \bar V}{\pa_v \bar V}\right) \pa_v T + 2 \pa_v\left(\frac{\delta \bar V}{\pa_v \bar V}\right) T.
  \label{Equation: T variation outside}
\end{align}
With this we can establish that
\begin{align}
  \delta\qty(\int_{\SegmentBar}\dd{v} \frac{\delta\bar V}{\partial_v\bar V} T) &= \int_{\SegmentBar}\dd{v} \frac{\delta\bar V}{\partial_v\bar V}\wedge \qty(\frac{\partial_v\delta\bar V}{\partial_v\bar V} T - \delta T)
  = \frac12\int_{\SegmentBar}\dd{v}\delta T\wedge \frac{\delta\bar V}{\partial_v\bar V}.
\end{align}

Let us now compute the analogous contributions coming from the anomalous counterterm in the symplectic form.

We look first at the contributions from $\SegmentBar$. From~\eqref{Equation: eta outside} we have
\begin{equation}
  \delta \eta|_{\SegmentBar} = \delta q_a + \partial_v \eta \qty( \frac{\delta\bar V}{\partial_v\bar V}) + \pa_v \qty( \frac{\delta\bar V}{\partial_v\bar V}),
\end{equation}
so that
\begin{align}
  \partial_v\delta \eta \wedge\delta \eta &= \partial_v\delta\eta\wedge \qty(\delta q_a +  \qty( \frac{\delta\bar V}{\partial_v\bar V}) \partial_v \eta + \pa_v\qty( \frac{\delta\bar V}{\partial_v\bar V}))\\
  &= \delta \qty( \frac12 (\pa_v \eta)^2 -\pa_v^2 \eta)\wedge\frac{\delta \bar V}{\partial_v\bar V} - \pa_v\left( \delta q_a \wedge\delta \eta + \frac{\delta\bar V}{\partial_v\bar V}\wedge\partial_v\delta \eta\right).
\end{align}
Note that the first term is 
$24\pi T_{\text{an}}\wedge\frac{\delta \bar V}{\partial_v\bar V}$. 
Using the falloff conditions at $\epsilon_a\infty$ to discard the boundary term there, we therefore have
\begin{equation}
  {\bm\Omega}_{\text{an}} = \frac{1}{48\pi}\int_{\Segment}\dd{v}\partial_v\delta\eta\wedge\delta\eta + {\bm\Omega}_{\text{an}}^1-{\bm\Omega}_{\text{an}}^0,
\end{equation}
where
\begin{equation}
  {\bm\Omega}_{\text{an} }^a = \frac{1}{48\pi}\int_{v_a}^{\epsilon_a\infty}\partial_v\delta \eta \wedge\delta \eta = \frac12\int_{v_a}^{\epsilon_a\infty}\delta T_{\text{an}}\wedge \frac{\delta\bar V}{\partial_v\bar V} - \frac{1}{48\pi}\qty(\delta v_a \wedge \delta \beta_a- \delta q_a\wedge (\delta\eta_a - \beta_a\delta v_a)).
\end{equation}
Here we have  used\footnote{We also have that
  $
[\pa_v\delta \eta](v_a)= \delta \beta_a - (\pa_v\beta)(v_a)\delta v_a $, and the $\delta v_a$ term conveniently disappears from the wedge product.}
\begin{equation}
  \delta \bar{V}(v_a)= -\pa_v\bar{V}(v_a) \delta v_a, \qquad
  [\delta \eta](v_a) = \delta\eta_a - \beta_a\delta v_a.
\end{equation}

Now we can focus on the anomalous contributions to the symplectic structure supported on $\mathcal{I}$. Using $\eta|_{\Segment}= \ln \pa_v V$ one finds
\begin{align}
  \int_{\mathcal{I}}\dd{v}\partial_v\delta \eta\wedge \delta \eta
  &= \int_{\mathcal{I}}\dd{v} \delta\qty(\beta \frac{\partial_v \delta V}{\partial_v V}) \\
  &= \int_{\mathcal{I}}\dd{v} \delta \qty(\partial_v \qty(\frac{\beta}{\partial_vV}\delta V)-\partial_v\qty(\frac{\beta}{\partial_vV})\delta V) \\
  &= -\int_{\mathcal{I}}\dd{v} \delta \partial_v\qty(\frac{\pa_v \eta}{\partial_vV})\wedge\delta V +\sum_a \epsilon_a( \beta_a\delta\eta_a-\delta\beta_a)\wedge\delta v_a.
\end{align}
Combining this with the above, the anomalous counterterm in the symplectic form can be written as
\begin{equation}
{\bm\Omega}_{\text{an}} = -\frac{1}{48\pi}\delta\qty(\int_{\Segment}\pa_v\qty(\frac{\pa_v \eta}{\partial_vV})\wedge \delta V) + \frac12\int_{\SegmentBar}\delta T_{\text{an}} \wedge\frac{\delta\bar V}{\partial_v \bar V} - \frac{1}{48\pi}\sum_a\epsilon_a\delta\bar\eta_a\wedge\qty(\delta\eta_a-\beta_a\delta v_a)
  \label{Equation: symplectic counterterm segment}
\end{equation}

One may now straightforwardly combine the tree-level symplectic form~\eqref{Equation: tree level symplectic form segment} with the anomalous contribution~\eqref{Equation: symplectic counterterm segment} to obtain~\eqref{Equation: segment reduced anomalous symplectic form decomposition}, as required.

Finally, let us confirm $\delta\bm\Omega_{\SegmentBar}=0$. Using
\begin{equation}
  \delta T_c|_{\SegmentBar} = \frac{\delta\bar V}{\partial_v \bar V}\partial_v T_c + 2T_c\partial_v\qty(\frac{\delta\bar V}{\partial_v \bar V}) - \frac{c\hbar}{24\pi}\partial_v^3\qty(\frac{\delta\bar V}{\partial_v \bar V}),
\end{equation}
we have
\begin{align}
  \delta\qty(\frac12\int_{\SegmentBar}\dd{v}\delta T_c\wedge\frac{\delta \bar V}{\partial_v \bar V}) &= \frac12\int_{\SegmentBar}\dd{v}\delta T_c\wedge\partial_v\qty(\frac{\delta\bar V}{\partial_v\bar V})\wedge\frac{\delta \bar V}{\partial_v \bar V}\\
  &=-\frac{c\hbar}{48\pi}\int_{\SegmentBar}\dd{v}\partial_v^3\qty(\frac{\delta\bar V}{\partial_v\bar V})\wedge\partial_v\qty(\frac{\delta\bar V}{\partial_v\bar V})\wedge\frac{\delta \bar V}{\partial_v \bar V}\\
  &=-\frac{c\hbar}{48\pi}\int_{\SegmentBar}\dd{v}\partial_v\qty(\partial_v^2\qty(\frac{\delta\bar V}{\partial_v\bar V})\wedge\partial_v\qty(\frac{\delta\bar V}{\partial_v\bar V})\wedge\frac{\delta \bar V}{\partial_v \bar V})\\
  &= -\frac{c\hbar}{48\pi}\sum_a\epsilon_a\delta\beta_a\wedge\delta\bar\eta_a\wedge\delta v_a.
\end{align}
Hence, $\bm\Omega_{\SegmentBar}$ is closed, as claimed.

\subsection{Phase space in dressed variables}
\label{Subsection: anomalous dressed phase space}

Let's write the segment symplectic structure $\bm\Omega_{\Segment}$ in terms of the embedding field $X$ and dressed fields $\tilde\tau,\tilde\varphi_i$. First, note that
\begin{align}
  \int_{\Segment}\dd{v}\tau_c\frac{\delta V}{\partial_v V} &= \int_{I}\dd{\mv} (\partial_vX)^2(\tau_c\circ X) \delta V\circ V^{-1} = -\int_{I}\dd{\mv} (\partial_vX)^2(\tau_c\circ X) \frac{\delta X}{\partial_\mv X}.
\end{align}
Using $\tau_c = \tau -\frac{c\hbar}{48\pi} \partial_v V\pa_v \left(\frac{\pa_v^2 V}{(\pa_vV)^2}\right)$ one may find the infinitesimal action of a diffeomorphism on $\tau_c$:
\begin{equation}
  \delta_f\tau_c = f\partial_v\tau_c+2\tau_c\partial_v f - \frac{c\hbar}{48\pi}\partial_v V\partial_v\qty\bigg(\frac{\partial_v^2f}{\partial_vV}).
  \label{Equation: Pi transformation infinitesimal}
\end{equation}
This action can be exponentiated to
\begin{equation}
  F\triangleright\tau_c = (\partial_v F)^2\tau_c \circ F- \frac{c\hbar}{48\pi}\qty[\partial_v\qty(\frac{\partial_v^2F}{\partial_vF})-\frac{\partial_v^2F}{\partial_vF}\frac{\partial_v^2(V\circ F)}{\partial_v(V\circ F)}].
  \label{Equation: Pi transformation}
\end{equation}
So one has
\begin{align}
  (\partial_\mv X)^2\tau_c\circ X = X\triangleright \tau_c + \frac{c\hbar}{48\pi}\partial_\mv\qty(\frac{\partial_\mv^2 X}{\partial_\mv X}).
\end{align}
Now $\tilde{\tau}_c
:= X\triangleright\tau_c=X\triangleright\tau=\tilde\tau$ (since $X\triangleright \eta=0$, see \eqref{etaaction}).
Thus,
\begin{equation}
  \int_{\Segment}\dd{v}\tau_c\frac{\delta V}{\partial_v V} = - \int_{I}\dd{\mv} \qty(\tilde\tau + \frac{c\hbar}{48\pi}\partial_\mv\qty(\frac{\partial_\mv^2 X}{\partial_\mv X}))  \frac{\delta X}{\pa_\mv X} .
\end{equation}
We also have (as in the tree-level case)
\begin{equation}
  \int_{\Segment}\dd{v}\partial_v\varphi_i \delta\varphi_i = \int_I\dd{\mv}\qty(\partial_\mv\tilde\varphi_i\delta\tilde\varphi_i -(\partial_\mv\tilde\varphi_i)^2\frac{\delta X}{\pa_\mv X}).
\end{equation}
Using the above in~\eqref{Equation: anomalous Omega I}, one finds
\begin{align}
  \bm\Omega_{\Segment}
  &= \int_I\dd{\mv}\delta\qty(-\qty(\tilde{T}+\frac{c\hbar}{48\pi}\partial_\mv\qty(\frac{\partial_\mv^2 X}{\partial_\mv X}))\frac{\delta X}{\pa_\mv X} + \sum_i\partial_\mv\tilde\varphi_i\delta\tilde\varphi_i),
  \label{Equation: fully dressed anomalous}
\end{align}
where
\begin{align}
  \tilde{T} = X\triangleright T
  =  \tilde\tau+\sum_i(\partial_\mv\tilde\varphi_i)^2
\end{align}
is the dressed stress tensor (note that its form is unchanged from the tree-level case).

\subsection{Virasoro Maurer-Cartan form}
\label{Subsection: Virasoro Maurer-Cartan}

Throughout the paper so far, we have made extensive use of the left and right Maurer-Cartan 1-forms for the diffeomorphism group
\be
\theta_{\text{diff}}= \delta X\circ X^{-1}=-\frac{\delta V}{\pa_v V},
\qquad
\tilde\theta_{\text{diff}}= \delta V\circ V^{-1}=-\frac{\delta X}{\pa_\mv X}.
\label{Equation: diff Maurer-Cartans}
\ee
Indeed, these played a central role in the symplectic form for the spin 0 degrees of freedom at tree-level. They appear similarly in the effective spin 0 symplectic form~\eqref{Equation: anomalous Omega I}. But in the effective theory, the diffeomorphism group is replaced by the Virasoro group, and this is suggestive that one should really be making use of the Maurer-Cartan forms for the Virasoro group, instead of~\eqref{Equation: diff Maurer-Cartans}.

It is interesting and useful to note that the effective spin 0 symplectic potential
\be
\bm\Theta^0_{\mathcal{I}}
=-
\int_I \rd \mv \left( \tilde{\tau} \frac{\delta X}{ \pa_\mv X}
  +\frac{c\hbar}{48\pi} \pa_\mv\left( \frac{\pa_\mv^2 X}{ \pa_\mv X} \right)
\frac{\delta X}{ \pa_\mv X} \right)
\label{Equation: effective spin 0 symplectic potential}
\ee
can indeed be understood in this way. In particular, it is the pairing of an element  $(\tilde{\tau}- c)$ in the dual of the Virasoro algebra, with the right-invariant Virasoro Maurer-Cartan form $\tilde\theta_{\text{Vir}}$.
Thus, the spin 0 symplectic form is a KKS symplectic form for the Virasoro group. This point of view will allow us to efficiently derive diffeomorphism charges in the effective theory.

Recall that for a group $G$ with Lie algebra $\mathfrak{g}$ and Lie bracket $[\cdot,\cdot]$, the Maurer-Cartan form is a $\mathfrak{g}$ valued 1-form on the group defined by $\tilde\theta=\delta g\circ g^{-1}$. It satisfies
\be
\delta \tilde\theta +\frac12 [\tilde\theta,\tilde\theta]=0.
\label{Equation: MC Flatness}
\ee
Given an element $\tilde\tau = \mathrm{Ad}^*_g \tau $ (we have been denoting this as $ g\triangleright \tau$) valued in the dual $\mathfrak{g}^*$ of the Lie algebra one has
\be
\delta \tilde\tau = \mathrm{ad}^*_{\tilde\theta}(\tilde{\tau}),
\ee
for variations that leave $\tau$ invariant. Here $\mathrm{Ad}^*, \mathrm{ad}^*$ are the coadjoint actions of $G,\mathfrak{g}$ respectively on $\mathfrak{g}^*$, i.e.\ the duals of the corresponding adjoint actions on $\mathfrak{g}$. Finally, recall that the KKS symplectic potential is simply given by the pairing $\langle \tilde{\tau}| \tilde\theta \rangle$.

In our case we consider as Lie algebra the Lie algebra of the diffeomorphism group of $I$; accordingly $\tilde\theta_{\text{diff}}=\delta V\circ V^{-1}=-\delta X/\partial_\mv X$ is valued in this algebra.\footnote{More precisely, $\delta V\circ V^{-1}$ is the $\mv$ component of $\tilde\theta_{\text{diff}}$, i.e.\ $\tilde\theta_{\text{diff}}=(\delta V\circ V^{-1})\partial_\mv$.}
It is right-invariant, i.e.\ it is invariant under reparametrizations (the right action $\delta_f V=f \pa_v V$), while it transforms as a vector field under reorientations or dressed reparametrizations (the left action $\delta_g V = g\circ V$):
\be
L_{\delta_f} \tilde\theta_{\text{diff}}=0, \qquad
L_{\delta_g} \tilde\theta_{\text{diff}}= -[g, \tilde\theta_{\text{diff}}].
\ee
It also satisfies $I_{\delta_g}\tilde\theta_{\text{diff}}= g$; in fact, this condition and the flatness property~\eqref{Equation: MC Flatness} suffice to uniquely determine the Maurer-Cartan form.

Dual elements of the Lie algebra are rank 2 tensor fields $\tilde\tau$, which are paired with vector fields $f\partial_\mv$ via integration:
\begin{equation}
  \braket{\tilde\tau}{f} = \int_I\dd{\mv}\tilde\tau f.
\end{equation}
One can readily recognize the KKS potential form $\braket*{\tilde\tau}{\tilde\theta_{\text{diff}}}$ as the tree-level spin 0 symplectic potential.

Now let us see how this extends to the Virasoro group. Elements of this group are a tuple $(X,\alpha)$ of a diffeomorphism $X\in \mathrm{Diff}(I)$ and a number $\alpha$ (the central element). The group composition law is
\begin{equation}
  (X,\alpha)\circ (X',\alpha') = \qty\bigg(X\circ X', \alpha+\alpha' + \frac{\hbar}{48\pi}\int_I\dd{\mv}\log(\partial_\mv (X\circ X'))\frac{\partial_\mv^2 X'}{\partial_\mv X'})  \label{Equation: Virasoro composition}
\end{equation}
(usually the group is defined over $\RR$ or $S^1$; here we define it on the interval $I$). The integral appearing on the right is known as the Bott-Thurston cocycle \cite{bott1977characteristic, oblak2017berry}. It is important to note that this satifies the cocycle identity\footnote{That is 
$B(F,G)= \int_I \ln\pa_{\mv}(F\circ G) \rd \ln \pa_{\mv} G  $ satisfies the cocycle identity
\be 
B(FG,H)+B(F,G)=B(F,GH)+ B(G,H).
\ee  for $F,G,H \in \mathrm{Diff}(I)$ when $\pa_{\mv}H(0)=\pa_{\mv}H(1)$ and similarly for $G$.}
under the conditions $X(a)=a$ and $\pa_{\mv}X(0)=\pa_{\mv}X(1)$. The inverse is
\begin{equation}
  (X,\alpha)^{-1} = \qty(X^{-1}, -\alpha) = \qty(V,-\alpha).
  \label{Equation: Virasoro inverse}
\end{equation}
Elements of the Lie algebra of the Virasoro group may be decomposed as the direct sum of a vector field $g\partial_\mv$ (corresponding to the diffeomorphism) and a number $k$ (corresponding to the central element); we denote such an element as $g+k$.
From~\eqref{Equation: Virasoro composition} one may derive the right-invariant Maurer-Cartan form (see also \cite{Alekseev:2022qrq})
\begin{align}
  \tilde\theta_{\text{Vir}}
  &= \delta(X,\alpha)^{-1}\circ (X,\alpha) = \tilde\theta_{\text{diff}} - \frac{\hbar}{48\pi} \int_I \rd \mv \pa_\mv\left(\frac{\pa_\mv^2 X}{\pa_\mv X}\right) \tilde\theta_{\text{diff}}  -\delta \alpha,
  \label{Equation: Virasoro Maurer Cartan}
\end{align}
where $\tilde\theta_{\text{diff}}=-\delta X/\partial_\mv X$.
One may also derive the Virasoro bracket
\be
[g+k,g'+k']_{\text{Vir}}
= [g,g'] - \frac{\hbar}{48\pi}\int_{I} \dd{\mv}(g \pa_\mv^3 g'-g'\pa_\mv^3 g),
\ee
where $[g,g']=g\partial_\mv g'-g'\partial_\mv g$ is the usual vector field Lie bracket, and the left action:
\begin{equation}
  \delta_{g+k} X = -g\partial_\mv X, \qquad \delta_{g+k}\alpha = - \frac{\hbar}{48\pi} \int\dd{\mv}g\partial_\mv\qty(\frac{\partial_\mv^2 X}{\partial_\mv X}) - k.
\end{equation}
One may confirm that $I_{\delta_{g+k}}\tilde\theta_{\text{Vir}}=g+k$.

To obtain these expressions, we had to integrate by parts and neglect boundary terms, which is only valid when the appropriate boundary conditions are satisfied. A proper treatment involves the edge modes, and we shall address this in more detail below.

Dual elements of the Virasoro algebra are the direct sum of a rank 2 tensor field $\tilde\tau$ and a number $c$ (which is the central charge); we denote elements of the dual as  $\tilde\tau - c$. The pairing is defined as
\begin{equation}
  \braket{\tilde\tau- c}{g +k} = \int_I\dd{\mv}\tilde\tau g -ck.
\end{equation}
Using~\eqref{Equation: Virasoro Maurer Cartan}, one finds
\begin{equation}
  \braket*{\tilde \tau - c}{\tilde\theta_{\text{Vir}}} = -
  \int_I \rd \mv \left( \tilde{\tau} \frac{\delta X}{ \pa_\mv X}
    +\frac{c\hbar}{48\pi} \pa_\mv\left( \frac{\pa_\mv^2 X}{ \pa_\mv X} \right)
  \frac{\delta X}{ \pa_\mv X} \right) + c\,\delta \alpha.
  \label{Equation: Virasoro symplectic potential}
\end{equation}
As promised above, this is a symplectic potential for the spin 0 degrees of freedom~\eqref{Equation: effective spin 0 symplectic potential} (since we are fixing the central charge, the extra term appearing here $c\,\delta \alpha$ is closed and so does not affect the symplectic form). 

One point of sublety is that the dressing field $X$ is not a diffeomorphism of $I$. It maps $I$ onto $\mathcal{I}$ where the image involves the edge modes $v_a$. To  properly address the role played by the edge modes in the Virasoro Maurer-Cartan form we need to evaluate $\delta\tilde\theta_{\text{Vir}}$. Using
\begin{align}
  \delta \left(\frac{ \pa_\mv^2 X}{ \pa_\mv X}\right)
  &= \frac{\pa_\mv^2 \delta X}{ \pa_\mv X}  - \frac{\pa_\mv^2 X}{ \pa_\mv X} \frac{\pa_\mv \delta X}{ \pa_\mv X} = \pa_\mv \left(\frac{\pa_\mv \delta X}{ \pa_\mv X}\right)= \pa_\mv^2 \left(\frac{ \delta X}{ \pa_\mv X}\right) + \pa_\mv\left( \left(\frac{ \delta X}{ \pa_\mv X}\right) \frac{\pa_\mv^2 X}{ \pa_\mv X} \right),
\end{align}
we have
\begin{align}\label{varchiv}
  \delta \left( \pa_\mv\left( \frac{\pa_\mv^2 X}{ \pa_\mv X} \right)
  \frac{\delta X}{ \pa_\mv X} \right) &= - \frac{\delta X}{\pa_\mv X } \wedge \pa_{\mv}^3 \left( \frac{\delta X}{\pa_\mv X }\right)
  - \pa_{\mv}\left(
  \frac{\pa_\mv^2 X}{ \pa_\mv X} \frac{\delta X}{\pa_\mv X } \wedge \pa_{\mv}\left(\frac{\delta X}{\pa_\mv X } \right) \right).
\end{align}
The final term yields a boundary contribution in
\begin{align}
  \delta\tilde\theta_{\text{Vir}} &= \delta\tilde\theta_{\text{diff}} + \frac{\hbar}{48\pi} \int_I\dd{\mv}\qty[\frac{\delta X}{\pa_\mv X } \wedge \pa_{\mv}^3 \left( \frac{\delta X}{\pa_\mv X }\right) + \pa_{\mv}\left(
  \frac{\pa_\mv^2 X}{ \pa_\mv X} \frac{\delta X}{\pa_\mv X } \wedge \pa_{\mv}\left(\frac{\delta X}{\pa_\mv X } \right) \right)]\\
  &= -\frac12[\tilde\theta_{\text{Vir}},\tilde\theta_{\text{Vir}}]_{\text{Vir}} + \frac\hbar{48\pi}\left[
  \frac{\pa_\mv^2 X}{ \pa_\mv X} \frac{\delta X}{\pa_\mv X } \wedge \pa_{\mv}\left(\frac{\delta X}{\pa_\mv X } \right) \right]_0^1
\end{align}
where we have used $\delta\tilde\theta_{\text{diff}}=-\frac12[\tilde\theta_{\text{diff}},\tilde\theta_{\text{diff}}]$. One may write this in terms of the edge modes by using $X(a)=v_a$ and
\be
\pa_\mv X
=e^{-\eta }\circ X,
\qquad
\frac{\pa_\mv^2 X}{\pa_\mv X}
=
-( e^{-\eta} \beta) \circ X,
\ee
which implies
\begin{align}
  \frac{\delta X}{\pa_\mv X}(a) &= e^{\eta_a} \delta v_a
  \qquad
  \pa_\mv\left(\frac{\delta X}{\pa_\mv X}\right)(a) = - \delta \eta_a + \beta_a \delta v_a.
\end{align}
One finds
\begin{equation}
  \delta\tilde\theta_{\text{Vir}} + \frac12[\tilde\theta_{\text{Vir}},\tilde\theta_{\text{Vir}}] = -\frac\hbar{48\pi} \sum_a\epsilon_a\beta_a\delta\eta_a\wedge\delta v_a.
\end{equation}
Thus, the form $\tilde\theta_{\text{Vir}}$ does not quite obey the zero-curvature condition, in the presence of the edge modes, when $\delta v_a$ and  $\delta \eta_a $ are both non vanishing. 

It seems likely that it is possible to describe these structures using a unified group that accounts for both the bulk Virasoro structure and the edge modes, and whose Maurer-Cartan form obeys the zero-curvature condition even in the presence of non-zero $\delta v_a,\delta\eta_a$. We hope to properly investigate this possibility in future work.

\subsection{Anomalous diffeomorphism actions}
\label{Subsection: anomalous actions}

In Section~\ref{Section: diffeomorphism actions} we described reparametrizations, reorientations and dressed reparametrizations in the tree-level theory. These diffeomorphism actions continue to play a role in the effective theory, but they are anomalous, acquiring central charges according to Table~\ref{Table: anomalous diffeomorphisms}.

In the effective theory, the importance of the dressed reparametrizations becomes more clear. They are generated by the dressed constraint $\tilde T$, and this is the constraint that should be imposed in the effective theory; see Subsection~\ref{Subsection: anomalous constraint} for more detail.

\subsubsection{Reparametrizations}

We have already demonstrated that, before reducing to the null ray segment, the generator of a reparametrization is the anomalous stress tensor, i.e.\ $I_{\delta_f}\bm\Omega^c=-\delta T^c_{f}$ where
\begin{equation}
  T^c_f = \int_{\Ray} \dd{v} f T^c.
\end{equation}
Moreover, we showed that these obey the algebra
\begin{equation}
  \pb*{T^c_f}{T
  ^c_{f'}}=-T^c_{[f,f']}+\frac{c\hbar}{48\pi}\int_\RR\dd{v}(f\partial_v^3f'-f'\partial_v^3f).
\end{equation}
After reducing to the segment, $T^c_f$ continues to be the generator, and it satisfies the same algebra. This is a consequence of the fact that we did not break any gauge symmetries when doing the reduction, and it can be checked by a direct evaluation using the segment symplectic structure.

The central charge is $c$, and recall that the reparametrizations form an antimorphism of the underlying Virasoro algebra.

\subsubsection{Reorientations}

Let us now consider how reorientations are modified in the effective theory. Recall that at tree-level reorientations are transformations parametrized by a vector field $g(\mv)\partial_\mv$ with $g(0)=g(1)=0$, and act as \eqref{reor1}
\begin{equation}
  \delta_g V = g\circ V, \qquad  \delta_g\varphi_i=0, \qquad \delta_g\eta_a = \partial_\mv g(a).
\end{equation}
The above is still true in the effective theory. They correspond to a change in the frame value which leaves the original fields invariant.

However, under a reorientation, at tree-level $\tilde\tau$ transformed as a rank 2 tensor, and the edge modes $\omega_a$ corresponding to the area element at the endpoints were invariant, while in the effective theory these objects transform anomalously. The appropriate anomalous transformation for $\tilde\tau$ is
\be
\delta_g \tilde{\tau}
= -g\pa_\mv \tilde{\tau} - 2 \pa_\mv g \tilde{\tau} - \frac{c\hbar}{24\pi} \pa_\mv^3 g,
\label{Equation: anomalous reorientation tilde tau}
\ee
i.e.\ it transforms as an anomalous stress tensor at central charge $-c$ (the overall minus sign in~\eqref{Equation: anomalous reorientation tilde tau} implies reorientations form a morphism of the Virasoro algebra). The edge modes $\omega_a$ will be required to satisfy
\begin{equation}
  \delta_g\omega_a = \frac{c\hbar}{48\pi}\partial_\mv g(a).
\end{equation}
This is such that the values of the \emph{anomalous} area element $\Omega_c$ at the endpoints of the segment are invariant under a reorientation:
\begin{equation}
  \delta_g\qty(\frac{\Omega_c(v_a)}{8\pi\GN}) = \delta_g\qty(\omega_a - \frac{c\hbar}{48\pi}\eta_a) = 0.
  \label{Equation: anomalous reorientation area edge mode}
\end{equation}
Finally we have that reorientations leave the external frame invariant $\delta_g \bar{V}=0$.

Let us now find the generator of reorientations in the effective theory.
To this end, it is simplest to write the symplectic form as
\begin{equation}
  \bm\Omega = \delta\braket*{\tilde\tau-c}{\tilde\theta_{\text{Vir}}} + \sum_a\epsilon_a\delta(\omega_a-\tfrac{c\hbar}{48\pi}\eta_a)\wedge\delta q_a + \delta\qty(\int_{\Segment}\dd{v}\sum_i\partial_v\varphi_i\delta\varphi_i) + \bm\Omega_{\SegmentBar}.
\end{equation}
When computing $I_{\delta_g}\bm\Omega$, only the first two terms here contribute. One has
\begin{align}
  \delta\braket*{\tilde\tau-c}{\tilde\theta_{\text{Vir}}}
  &= \braket*{\delta\tilde\tau}{\tilde\theta_{\text{Vir}}} + \braket*{\tilde\tau-c}{\delta\tilde\theta_{\text{Vir}}} \\
  &= \braket*{\delta\tilde\tau}{\tilde\theta_{\text{Vir}}} -\frac12\braket*{\tilde\tau-c}{[\tilde\theta_{\text{Vir}},\tilde\theta_{\text{Vir}}]_{\text{Vir}}} - \frac{c\hbar}{48\pi} \sum_a\epsilon_a\beta_a\delta\eta_a\wedge\delta v_a.
\end{align}
Using $I_{\delta_g}\tilde\theta_{\text{Vir}}=g$ and the definition of the Virasoro bracket one then finds
\begin{align}
  I_{\delta_g}\delta\braket*{\tilde\tau-c}{\tilde\theta_{\text{Vir}}}
  &= -\delta\left(\int_I \rd \mv   \tilde{\tau} g\right) -\frac{c\hbar}{48\pi}\sum_a \epsilon_a   \pa_{\mv}g(a) \beta_a \delta v_a\\
    &\hspace{2em}+
  \int_I \rd \mv \bigg( \delta_g \tilde{\tau}  \tilde\theta_{\text{diff}}
    - \tilde{\tau} \left[ g, \tilde\theta_{\text{diff}} \right]
  -\frac{c\hbar}{48\pi} \left[ g \pa_{\mv}^3 \tilde\theta_{\text{diff}}  - \tilde\theta_{\text{diff}}   \pa_{\mv}^3 g\right] \bigg) 
\end{align}
With~\eqref{Equation: anomalous reorientation tilde tau},
the final integral reduces to a boundary term:
\begin{align}
  I_{\delta_g} \delta\braket*{\tilde\tau-c}{\tilde\theta_{\text{Vir}}}
  &= -\delta\left(\int_I \rd \mv   \tilde{\tau} g\right) -\frac{c\hbar}{48\pi}\sum_a \epsilon_a   \pa_{\mv}g(a) \beta_a \delta v_a\\
  &\hspace{2em}-\left[ g \tilde{\tau} \tilde\theta_{\text{diff}}  + \frac{c\hbar}{48\pi}\left( \pa_v^2 g \tilde\theta_{\text{diff}} + g \pa_\mv^2\tilde\theta_{\text{diff}} - \pa_\mv g \pa_\mv\tilde\theta_{\text{diff}}\right) \right]_0^1 \\
  &= -\delta\left(\int_I \rd \mv   \tilde{\tau} g\right) + \frac{c\hbar}{48\pi}\sum_a \epsilon_a
  \left( e^{\eta_a}\pa_{\mv}^2 g(a) \delta v_a - \pa_\mv g(a) \delta \eta_a \right)\\
  &= -\delta\left(\int_I \rd \mv   \tilde{\tau} g + \frac{c\hbar}{48\pi}\sum_a \epsilon_a \pa_\mv g(a)  \eta_a \right) + \frac{c\hbar}{48\pi}\sum_a \epsilon_a
   e^{\eta_a}\pa_{\mv}^2 g(a) \delta v_a .
  \label{Equation: reorientation in KKS anomalous}
\end{align}
To get the second equality, we used $g(a)=0$ and
\begin{equation}
    \partial_\mv\tilde\theta_{\text{diff}}(a) = -\left.\partial_\mv\qty(\frac{\delta X}{\partial_\mv X})\right|_a = \delta\eta_a - \beta_a\delta v_a.
\end{equation}
Using~\eqref{Equation: anomalous reorientation area edge mode}, we also have
\begin{equation}
I_{\delta_g}\left(\sum_a\epsilon_a\delta(\omega_a-\tfrac{c\hbar}{48\pi}\eta_a)\wedge\delta q_a \right)= -\delta\qty(\sum_a\epsilon_a\partial_\mv g(a)(\omega_a-\tfrac{c\hbar}{48\pi}\eta_a)),
\end{equation}
so in total one finds
\begin{equation}
I_{\delta_g}\bm\Omega = -\delta\left(\int_I \rd \mv   \tilde{\tau} g + \sum_a \epsilon_a \pa_\mv g(a)  \omega_a \right) + \frac{c\hbar}{48\pi}\sum_a \epsilon_a
 e^{\eta_a}\pa_{\mv}^2 g(a) \delta v_a.
\label{Equation: reorientations integrable}
\end{equation}
The last term is not integrable in general. Indeed, reorientations are integrable only if we impose the additional condition
\be
\pa_\mv^2 g(a)=0.
\ee
Note that this condition may be understood as coming from the continuity of $\beta$ at $v_a$, which is an implicit assumption we have been making in this paper. Indeed, $\delta_g\bar V=0$ implies $\delta_g\beta_a=0$, but
\begin{equation}
\delta_g\beta_a = \left.\partial_v \frac{\partial_v\delta_g V}{\partial_v V}\right|_{v_a} = e^{\eta_a}\partial_\mv^2 g(a).
\end{equation}
Therefore, we must have $\partial_\mv^2 g(a)=0$.

One then finds $I_{\delta_g}\bm\Omega = -\delta Q_g$, where
\begin{equation}
Q_g = \int_0^1\dd{\mv}g\tilde\tau +\sum_a\epsilon_a\partial_\mv g(a)\omega_a = \int_0^1\partial_\mv^2 g \tilde\Omega \dd{\mv}.
\end{equation}
Thus, the reorientation charges are the same in the effective theory as in the tree-level theory. However, the reorientation algebra is modified:
\begin{align}
\{Q_{g},Q_{g'}\}= \delta_{g}Q_{g'} &= -\int_{\Segment} g'\qty(g\partial_\mv\tilde\tau+2\tilde\tau\partial_\mv g + \frac{c\hbar}{24\pi}\partial_\mv^3 g) - \frac{c\hbar}{48\pi}\sum_a\epsilon_a\partial_\mv g(a) \partial_\mv g'(a)\\
&= \int_{\Segment} (g\partial_{\mv}g'-g'\partial_\mv g)\tilde\tau - \frac{c\hbar}{48\pi}\int_{\Segment} \dd{v}( g'\partial_\mv^3 g - g\partial_\mv^3g') \\
&= Q_{[g,g']} + \frac{c\hbar}{48\pi}\int_{\Segment} \dd{v}( g\partial_\mv^3 g' - g'\partial_\mv^3g)
\end{align}
So we have a Virasoro algebra at central charge $-c$.

For completeness, let's compute how the reorientation acts on the dressed area. Using~\eqref{area}, we have
\begin{align}
\frac{\delta_g\tilde\Omega(\mv)}{8\pi\GN} &= \mv\delta_g\omega_1 + (1-\mv)\delta_g\omega_0 + \int_0^1\dd{\mv'}\Greens{\mv}{\mv'}\delta_g\tilde\tau(\mv')\\
&=\int_0^1\dd{\mv'}\Greens{\mv}{\mv'}(g\partial_{\mv'}\tilde\tau+2\tilde\tau\partial_{\mv'}g)(\mv') \\
&\hspace{6em}+ \frac{c\hbar}{48\pi}\qty(\mv \partial_\mv g(1) + (1-\mv)\partial_\mv g(0) +2\int_0^1\dd{\mv'}\Greens{\mv}{\mv'}\partial_{\mv'}^3 g)\\
&=\int_0^1\dd{\mv'}\Greens{\mv}{\mv'}(g\partial_{\mv'}\tilde\tau+2\tilde\tau\partial_{\mv'}g)(\mv') + \frac{c\hbar}{48\pi}\qty(2\partial_\mv g(\mv)-\mv \partial_\mv g(1) - (1-\mv)\partial_\mv g(0) ).
\end{align}
We have integrated by parts. The first term is the tree-level result, and the second term is the effective correction.\footnote{It is interesting to note that the term proportional to $\partial_\mv g(v)$ appearing here is similar to the $\partial_vf$ term appearing in the transformation law for the anomalous area element~\eqref{Equation: Omega c transformation}.}

\begin{table}
  \centering
  \begin{tabular}{lccc}\toprule
    & \small
    \begin{tabular}{@{}c@{}}Central \\ charge
    \end{tabular}  & \small Generator & \small Algebra \\  \midrule
    {\small Reparametrization} & $c$ &$T^c_f = \int_{\Ray}\dd{v}fT^c$ & $\pb*{T^c_f}{T
    ^c_{f'}}=-T^c_{[f,f']}+\tfrac{c\hbar}{48\pi}\int_\RR\dd{v}(f\partial_v^3f'-f'\partial_v^3f)$\\\addlinespace
    {\small Reorientation} & $-c$ & $Q_g = \int_{I}\dd{\mv}g\tilde\tau$ & $\pb{Q_g}{Q_{g'}}=Q_{[g,g']}+\tfrac{c\hbar}{48\pi}\int_{I}\dd{v}(f\partial_v^3f'-f'\partial_v^3f)$\\\addlinespace
    {\small
      \begin{tabular}{@{}l@{}}Dressed \\ reparametrization
    \end{tabular}} & $-c$ &$\tilde T_{\tilde f} = \int_{I}\dd{\mv}\tilde f \tilde T$ & $\pb*{\tilde T_{\tilde f}}{\tilde T_{\tilde f'}}=\tilde T_{[\tilde f,\tilde f']}+\tfrac{c\hbar}{48\pi}\int_{I}\dd{v}(f\partial_v^3f'-f'\partial_v^3f)$
  \end{tabular}
  \caption{The effective theory inherits the three diffeomorphism actions from the tree-level theory, with some modifications: the generator of reparametrizations is changed from $T$ to $T^c$, and each diffeomorphism algebra develops anomalies.}
  \label{Table: anomalous diffeomorphisms}
\end{table}

\subsubsection{Dressed reparametrizations}

In the effective theory, the action of a dressed reparametrization is given by
\begin{equation}
\delta_{\tilde f} V = \tilde f\circ V, \qquad  \delta_{\tilde f}\tilde\varphi_i=0, \qquad \delta_{\tilde f}\bar V = 0, \qquad \delta_{\tilde f} q_a =0,
\end{equation}
which is the same as in the tree-level theory, and
\begin{equation}
\delta_{\tilde f} \tilde T = -\tilde f\partial_\mv\tilde T - 2\tilde T \partial_\mv \tilde f - \frac{c\hbar}{24\pi}\partial_\mv^3 \tilde f, \qquad \delta_{\tilde f}\qty(\omega_a-\frac{c\hbar}{48\pi}\eta_a)=0,
\label{Equation: anomalous dressed reparametrization}
\end{equation}
which is deformed from the tree-level theory.

For the same reason as with reorientations, the continuity of $\beta$ at $v=v_a$ implies that we must have $\partial_v^2\tilde{f}(a)=0$. Here we additionally have the condition that $\partial_v\tilde f(a)=0$ from $\delta_{\tilde{f}}q_a=0$.

The generator of this transformation is the same as in the tree-level theory: the dressed stress tensor
\begin{equation}
\tilde T = \tilde \tau + \sum_i(\partial_\mv\tilde\varphi_i)^2.
\end{equation}
Indeed, we will show below that
\begin{equation}
I_{\delta_{\tilde f}}\bm\Omega = -\delta \tilde T_{\tilde f}, \qq{where} \tilde T_{\tilde f} = \int_I\dd{\mv}\tilde f \tilde T.
\label{Equation: Equation: effective dressed reparametrization charge}
\end{equation}
However, the algebra is deformed; using~\eqref{Equation: anomalous dressed reparametrization} leads to
\begin{align}
\{\tilde T_{\tilde{f}},\tilde T_{\tilde{f}'}\}= \delta_{\tilde f}\tilde T_{\tilde f'} &= -\int_{I}\dd{\mv} \tilde f'\qty(\tilde f\partial_\mv\tilde T+2\tilde T\partial_\mv \tilde f + \frac{c\hbar}{24\pi}\partial_\mv^3 \tilde f) \\
&= \int_{I} \dd{\mv}(\tilde f\partial_{\mv}\tilde f'-\tilde f'\partial_\mv \tilde f)\tilde\tau - \frac{c\hbar}{48\pi}\int_{\Segment} \dd{\mv}( \tilde f'\partial_\mv^3 \tilde f - \tilde f\partial_\mv^3\tilde f') \\
&= \tilde T_{[\tilde f,\tilde f']} + \frac{c\hbar}{48\pi}\int_{I} \dd{\mv}( \tilde f\partial_\mv^3 \tilde f' - \tilde f'\partial_\mv^3\tilde f)
\end{align}
This is a Virasoro representation at central charge $-c$.

To show~\eqref{Equation: Equation: effective dressed reparametrization charge} is is useful to write the symplectic form as
\begin{equation}
\bm\Omega = \delta\braket*{\tilde T-c}{\tilde\theta_{\text{Vir}}} + \sum_a\epsilon_a\delta(\omega_a-\tfrac{c\hbar}{48\pi}\eta_a)\wedge\delta q_a + \delta\qty(\int_{I}\dd{\mv}\sum_i\partial_v\tilde\varphi_i\delta\tilde\varphi_i) + \bm\Omega_{\SegmentBar},
\end{equation}
which follows from~\eqref{Equation: fully dressed anomalous}.
One then has
\begin{align}
I_{\delta_{\tilde f}}\bm\Omega &= I_{\delta_{\tilde f}}\delta\braket*{\tilde T-c}{\tilde\theta_{\text{Vir}}}\\
&= -\delta\left(\int_I \rd \mv   \tilde{T} \tilde f - \frac{c\hbar}{48\pi}\sum_a \epsilon_a \pa_\mv \tilde f(a)  \eta_a \right) + \frac{c\hbar}{48\pi}\sum_a \epsilon_a
\left( e^{\eta_a}\pa_{\mv}^2 \tilde f(a) \delta v_a \right),
\end{align}
where the second line follows from the same derivation that led to~\eqref{Equation: reorientation in KKS anomalous}, with the substitutions $\tilde \tau\to \tilde T$, $g\to\tilde f$. Using $\partial_\mv \tilde f(a)=\partial_\mv^2\tilde f(a)=0$ then gives~\eqref{Equation: Equation: effective dressed reparametrization charge} as required.

The properties of the different diffeomorphism symmetries are summarized in Table~\ref{Table: anomalous diffeomorphisms}.

\subsection{Imposing the anomalous constraint}
\label{Subsection: anomalous constraint}

Let us now give some brief comments on what it means to impose gauge-invariance in the effective theory, i.e.\ in the presence of the anomaly. The key point is that although the notion of gauge-invariant observable is unchanged, it is no longer directly possible to impose the constraint on states within the effective classical description. At tree-level one sets $\tilde T=0$, but in the effective theory, the best one can do is to allow $\tilde T$ to take values in a minimal coadjoint orbit of the Virasoro group. Therefore, there are extra gauge-invariant degrees of freedom in the effective classical theory, characterized by this orbit. The orbit degrees of freedom decouple from the gauge-invariant dressed radiative fields, $\pb{\tilde T}{\tilde \varphi_i}=0$.\footnote{Of course, they do not decouple from the reorientation charge $\pb{\tilde T}{\tilde\tau}\ne 0$.}

Nevertheless, once one performs a quantization of the effective theory, an appropriate choice of $c$ allows for cancellation between the classical and quantum anomalies, thus removing these extra degrees of freedom, and rendering the entire procedure self-consistent~\cite{QuantumAreaTime}.

In general, gauge-invariance is imposed at the level of observables by restricting to those which are unchanged by gauge transformations, as discussed previously. At tree-level, this means observables $\mathcal{O}$ satisfying
\begin{equation}
\delta_f\mathcal{O} = \pb{T_f}{\mathcal{O}} = 0.
\label{Equation: tree-level gauge invariant}
\end{equation}
This is unchanged in the effective theory:
\begin{equation}
\delta_f\mathcal{O} = \pb{T^c_f}{\mathcal{O}} = 0.
\label{Equation: effective gauge invariant}
\end{equation}
The condition~\eqref{Equation: effective gauge invariant} appears at first to differ from~\eqref{Equation: tree-level gauge invariant} in that it uses the deformed stress tensor $T^c$ instead of $T$ (and the Poisson bracket of the effective theory instead of the Poisson bracket of the tree-level theory). However, both conditions reduce to $\delta_f\mathcal{O}=0$, and $\delta_f$ is the same in both theories. The central nature of the anomaly is crucial here, as it means there are no additional conditions coming from an application of the Jacobi identity:
\begin{equation}
\pb{T^c_f}{\mathcal{O}} = 0 \implies \pb{T^c_{[f,f']}}{\mathcal{O}} = \pb{T^c_{f'}}{\pb{T^c_{f}}{\mathcal{O}}} - \pb{T^c_{f}}{\pb{T^c_{f'}}{\mathcal{O}}} = 0.
\end{equation}
So the set of gauge-invariant observables is unaffected by the anomaly -- before imposing the constraint.

At tree-level, we impose the constraint by restricting to states satisfying $T=0$, and this constraint is first-class, meaning it is preserved by gauge transformations. But in the effective theory, the deformed constraint $T^c$ would be second-class, meaning gauge transformations no longer preserve the anomalous Raychaudhuri equation $T^c=0$:
\begin{equation}
\delta_f T^c \, \stackrel{T^c=0}{=}\,  \frac{c\hbar}{24\pi}\partial_v^3 f,
\end{equation}
(except in the special case where $\partial_v^3f=0$, which corresponds to an $\SLTR$ subgroup of vacuum-preserving diffeomorphisms). In this way, it is no longer consistent to impose the constraint $T^c=0$ while preserving complete gauge redundancy. It is also not appropriate to just revert to the tree-level constraint $T=0$, because the tree-level stress tensor $T$ does not form a closed algebra with respect to the Poisson brackets of the effective theory, and in any case is not the generator of gauge transformations.

In the tree-level theory setting $T=0$ eliminates unphysical degrees of freedom; the inability to set $T^c=0$ in the effective theory means these extra degrees of freedom are still there. They are measured by the value of the gauge-invariant stress tensor $\tilde T$, which takes values in a Virasoro coadjoint orbit. One cannot try to additionally impose $\tilde T=0$, because $\tilde T$ does not commute with other gauge-invariant observables (including itself, because of the anomaly). Hence, measuring other gauge-invariant observables would typically create non-trivial configurations for $\tilde T$.

As we shall see in~\cite{QuantumAreaTime}, in the quantum theory the puzzle is easily resolved. One may in fact cancel the dressed reparametrization anomaly, and thus remove the extra degrees of freedom. Doing so requires one to still have an anomaly for ordinary reparametrizations and reorientations. This is unlike the classical description given here, where if the anomaly vanishes for one of these diffeomorphism representations then it must also vanish for the others (because in all cases the central charge is proportional to $c$). In the quantum case one cancels the physical anomaly of $\tilde{T}$ without canceling the anomalies of $T$ and $Q$.

\section{Conclusion}
\label{Section: Conclusion}
In this paper, we have presented a classical analysis of localization, dressing and diffeomorphism anomalies along a gravitational null ray. 

The first main result we achieved is a localization of the gravitational degrees of freedom on a null ray segment. We showed that this was made possible by the inclusion of edge-mode variables $v_a$ that represent the segment endpoints' locations. We can understand these edge modes in two ways: on the one hand, from an external perspective, they represent the influence of the segment's complement on the choice of frame. On the other hand, they can be understood internally as the degrees of freedom that enable us to convert a gauge-fixing condition $\beta =0$ into a frame-fixing condition $V=\mv$, and thus allow us to construct an internal frame. The gauge-fixing condition is a second-order differential equation on the frame, and the edge modes are the additional degrees of freedom required to construct it. With the edge modes, we can convert the gravitational canonical pair of area and surface tension $(\Omega,\beta)$ into the alternative canonical pair of energy and time $(\tau,V)$. This allows us to construct localized gauge invariant observables that \emph{commute} with observables localized on the complement.

In the second part of the paper, we introduced a deformation of the Raychaudhuri constraint and null ray symplectic structure that allows us to include the quantum anomaly in the classical phase space. This led to an effective description of quantum-gravitational fluctuations compatible with the quantum anomaly. 

We identified three important diffeomorphism actions (reparametrizations, reorientations, and dressed reparametrizations), and computed their central charges.
\begin{equation}
c_{T} = c,
\qquad
c_{\tilde\tau} = -c,
\qquad
c_{\tilde T} = -c.
\end{equation}

Both the localization and the choice of anomalous stress tensor were based on the dressing field which trivializes $\beta$, see~\eqref{Equation: tilde beta}. 

From our perspective, the dressing time possesses three distinctive features that make it favored over all other choices: First, the commutator of the dressing fields at different times Poisson-commutes. At the quantum level, this means that the dressing time is the most efficient reference frame in the sense that it minimizes its Heisenberg uncertainty. Second, it is such that the classical dressed constraint satisfies a generalization of the Wald-Zoupas criteria \cite{Wald2000} as explained in \cite{Odak:2023pga}. In particular, this means that the dressed constraint equation simply reads $\pa_{\mv}^2\tilde{\Omega}= - \sum_i (\pa_{\mv} \tilde\varphi_i)^2$ with no additional spin $0$ energy involved. This implies that the expansion is constant when there is no radiation and, moreover, that the area element in dressing time satisfies an analog of the Generalized Second Law (GSL)~\cite{Ciambelli:2023mir}.

It is still an interesting question to wonder what would happen if we considered a different dressing for the localization and/or the anomalous extension? For example, one can define a different dressing field $ V_\rho$ such that
\begin{equation}
\beta + \rho = \frac{\partial_v^2V_\rho}{\partial_vV_\rho},
\end{equation}
where $\rho$ is any field that transforms like a 1-form, i.e.
$
\delta_f\rho = \partial_v(f\rho).
$
Then $V_\rho$ defined in this way transforms appropriately for a dressing field:
\begin{equation}
F\triangleright V_\rho = V_\rho\circ F.
\end{equation}
So one can use it to dress other fields and obtain gauge-invariant observables. For example, by picking
\begin{equation}
\rho = \rho_{\text{aff}} := -4\pi\GN\frac1{\sqrt{\Omega}}\partial_v\qty\bigg(\frac1{\sqrt{\Omega}}\sum_i\varphi_i^2) + \frac12\theta,
\end{equation}
where $\theta$ is the expansion, one would obtain the affine time $V_{\rho}=V_{\text{aff}}$.\footnote{Recall~\eqref{Equation: half-densitisation} and $\mu=\kappa-\frac12\theta$ with $\kappa$ the inaffinity.}

Such a dressing field would be non-commuting and would thus considerably complicate the quantization achievable with the dressing time choice~\cite{QuantumAreaTime}.

One could also wonder what would happen if the anomalous extension were chosen by adding the Schwarzian derivative of a time coordinate that is different from the localization time.
Again, this would considerably complicate the construction of edge modes, and of the reorientation charges. It would also imply that the dressed constraints are different from the reduced constraints. So, this would be neither natural nor simple. 

The choice we have made here seems by far to be the most natural. But it would be interesting to understand whether this choice is \emph{uniquely} determined by physical criteria such as the ones outlined above: Wald-Zoupas, GSL, and minimal fluctuations.

A related issue concerns the interactions between the frame and the system. In this paper, we used a decoupling regime, in which the spin 0 degrees of freedom (the frame) are decoupled from the radiative degrees of freedom (the system). It will be important in the full theory to turn the interactions back on. The kinematical degrees of freedom will be the same, but the structure of the gauge-invariant degrees of freedom can be fundamentally altered by such interactions~\cite{Dittrich_2017,dittrich2015chaosdiracobservablesconstraint,Paiva_2022}. Moreover, one needs to take into account the renormalization of these interactions in the quantum theory, which can affect the structure of the effective theory, including the exact forms of the anomaly counterterms. Understanding this in more detail is another task for future work.

In this paper, we have focused on a single null ray and addressed the group of diffeomorphisms that act along it. To address the physics of a full null surface, one has to find a way to consistently glue all the null rays together and account for diffeomorphisms that act transversally to the rays. This will introduce new layers of complication. For example, the tree-level Raychaudhuri constraint transforms as a density with respect to transverse diffeomorphisms, and one would expect this to be true at the effective level also, but na\"ively this does not seem like it will be the case, at least with the effective theory described in this paper. To illustrate the issue, we can examine the anomalous area element
\begin{equation}
\Omega_c = \Omega - \frac{c\GN\hbar}{6}\eta.
\end{equation}
The tree-level area element $\Omega$ transforms like a density, but the anomalous term $\eta$ only transforms like a scalar, with respect to transverse diffeomorphisms. One way to cure this mismatch is to promote the central charge to a density depending on the transverse directions; this may be related to the `embadons' of~\cite{Ciambelli_2024}. Transverse diffeomorphisms will in principle, also be subject to anomalies, which should be accounted for in the effective theory by appropriate counterterms, and it seems plausible that a central charge density may be more natural in such a framework. We are very interested in resolving these questions in the future.

Throughout the paper, we extensively used the Maurer-Cartan forms of both the bulk diffeomorphism and Virasoro groups for the dressing time and the corner symmetry group for the edge modes. However, we did not have a unified picture of a single Maurer-Cartan form governing both the dressing time and the edge modes. We believe such a structure is possible and should be physically illuminating, and hope to discover it in later work. A related point is that the corner symmetry group on the null ray segment was found to reduce to the affine group -- but in the context of the Virasoro group, a more relevant structure ought to be based on $\SLTR$. It seems likely that various issues can be tackled by extending the edge modes such that they transform under an $\SLTR$ corner symmetry group. For example, recall that we found that reorientations $\delta_g$ in the effective theory are only integrable when $\partial_\mv^2g$ vanishes at the endpoints, see~\eqref{Equation: reorientations integrable}. By including edge modes transforming under $\SLTR$, one should be able to cancel the final term in that equation, and hence allow for integrable reorientations for which $\partial_\mv^2g(a)$ is non-zero.

There are also some remaining questions about the physical nature of the edge modes. We have mostly considered smooth diffeomorphisms in this paper, but as alluded to briefly in Section~\ref{Subsection: reparametrizations}, the presence of the edge modes at the endpoints of a segment allows one to break this condition, so long as one treats the edge modes as being physical sources of energy-momentum, as is done for example in~\cite{CLPW,kirklin2024generalisedsecondlawsemiclassical}. This perspective is compatible with the original proposal of~\cite{Donnelly:2016auv}, where the edge modes were designed to allow spacetime to be cut open. It would be important to understand if that option is available at the quantum level.

In our case, we have  remained fairly agnostic about the origin of the edge modes. Within our framework, they are part of the external dressing field $\bar{V}$, but we did not say where that field came from. It would be interesting to dig deeper into this question and investigate the relationship between the picture of  edge modes as additional internal  degrees of freedom and  the one where edge modes are made out of the fields that already existed in the theory (for instance, it could be a dressing time for the exterior), as in for example~\cite{Giddings_2019,Donnelly:2016rvo,Donnelly2016}. This question should be essential for the quantization of open regions.

Finally, the original motivation for this work was the \emph{quantization} of the dressing time reference frame. As already mentioned in the introduction, upon quantization, the central charges of the three diffeomorphism representations are shifted by different amounts. By appropriately tuning the classical central charge $c$, we can cancel the anomaly in the dressed reparametrization algebra at the quantum level, and thereby consistently quantize the classical Raychaudhuri equation. This will be described in full detail in~\cite{QuantumAreaTime}.

\newpage
\phantomsection
\addcontentsline{toc}{section}{\numberline{}Acknowledgements}
\section*{Acknowledgements}
We thank Rodrigo Andrade e Silva, Luca Ciambelli, Julian De Vuyst, Thomas Faulkner, Philipp H\"ohn, Max Lock, Luca Marchetti, Gon\c{c}alo Araujo Regado, Francesco Sartini, Bilyana Tomova, Tom Wetzstein, and Kona for helpful and stimulating discussions. We are grateful for the hospitality of the Okinawa Institute of Science and Technology, where part of this work was completed. This work was
supported by the Simons Collaboration on Celestial Holography.
This project was also made possible through the support of the ID\# 62312 grant from the John Templeton Foundation, as part of the \href{https://www.templeton.org/grant/the-quantum-information-structure-of-spacetime-qiss-second-phase}{\textit{`The Quantum Information Structure of Spacetime'} Project (QISS)}.~The opinions expressed in this project/publication are those of the author(s) and do not necessarily reflect the views of the John Templeton Foundation. Research at Perimeter Institute is supported in part by the Government of Canada through the Department of Innovation, Science and Economic Development and by the Province of Ontario through the Ministry of Colleges and Universities.

\printbibliography

\end{document}